\documentclass[fleqn,usenatbib]{mnras}
\usepackage[T1]{fontenc}
\usepackage{ae,aecompl}
\usepackage[latin9]{inputenc}
\usepackage{newtxtext,newtxmath}
\usepackage{mathrsfs}
\usepackage{amssymb}
\usepackage{graphicx}
\usepackage{wasysym}

\title[Shaping of the inner Oort cloud by Planet Nine]{Shaping of the inner Oort cloud by Planet Nine}

\author[Erez Michaely and Abraham Loeb]{
Erez Michaely and $^{1}$\thanks{E-mail: erezmichaely@gmail.com },
Abraham Loeb$^{2}$
\\
$^{1}$Physics Department, Technion - Israel Institute of Technology, Haifa
3200004, Israel\\
$^{2}$Department of Astronomy, Harvard University, 60 Garden St., Cambridge,
MA 02138, USA\\
}

\date{Accepted XXX. Received YYY; in original form ZZZ}

\pubyear{2017}

\begin{document}
\label{firstpage}
\pagerange{\pageref{firstpage}--\pageref{lastpage}}
\maketitle
\begin{abstract}
We present a numerical simulation of the dynamical interaction between
the proposed Planet Nine and a debris disk around the Sun for $4{\rm Gyr}$, accounting for the secular perturbation
of the four giant planets in two scenarios: (a) an initially thin circular disk
around the Sun (b) inclined and eccentric disk. We show, in both scenarios, that Planet Nine governs the dynamics
in between $1000{\rm -5000{\rm AU}}$ and forms spherical structure
in the inner part ($\sim1000{\rm AU}$) and inclined disk. This structure is the outcome of
mean motion resonances and secular interaction with Planet Nine. We
compare the morphology of this structure with the outcome from a fly-by
encounter of a star with the debris disk and show distinct differences
between the two cases. We predict that this structure serves as
a source of comets and calculate the resulting comet production rate
to be detectable.
\end{abstract}


\section{Introduction}
\label{sec:Introduction}

The structure of the outer solar system is complex and interesting.
\citet{Oort1950} explained the isotropic distribution of long-period
comets, by conjecturing with the existence of a Sun-centred
spherical cloud with inner semi-major axis (sma) of $2\times10^{4}{\rm AU}$
and an outer sma of $2\times10^{5}{\rm AU}$. Later work by \citet{Marsden1978}
extended the inner sma to $1\times10^{4}{\rm AU}$. Oort showed that
the comet reservoir can easily be perturbed by frequent distant stellar
encounter and the Galactic tides. These perturbations can drive the
objects in the Oort cloud to the inner-solar system with eccentricities
close to unity, yielding the constant rate of Sun grazing comets.

\citet{Hills1981} suggested that the inner boundary of the Oort cloud
reflects an observational bias, and the actual inner boundary is closer
than $2\times10^{4}{\rm AU}$. He found that the critical sma $a_{c}$
that satisfies both the rate of stellar encounters and a typical comet
lifetime is $a_{c}=2\times10^{4}{\rm AU}$. In this model, the inner
Oort cloud (``Hills cloud'') continues inward of $2\times10^{4}{\rm AU}$
but stellar encounters that on this scale occur on a longer timescale
than the typical lifetime of comets. Therefore, on these scales there
should be a burst of comet showers after each stellar encounter.

\citet{Duncan1987} studied numerically how the Oort and Hills clouds
were formed. They focused on scattering by the planets to high sma
and the interaction with the Galactic tides and stellar perturbations
and found that the inner part of the Hills cloud is $\sim3\times10^{3}{\rm AU}$. 

Recently \citet{Batygin2016} reported intriguing evidence for the
existence of a ninth planet in our solar system \citep{Batygin2016,Trujillo2014}.
The proposed Planet Nine has a mass of $m_{9}=10m_{\oplus}$, sma
of $a_{9}=700{\rm AU}$, eccentricity $e_{9}=0.6$, inclination to
the ecliptic, $i_{9}=30^{\circ}$, and argument of the periapsis,
$\omega_{9}=130^{\circ}$. Several other follow up studies supported
the existence of Planet Nine by different methods \citep{Fienga2016,Holman2016,Bailey2016,Gomes2017,Lai2016}.
Additionally, \citet{Li2016} calculated the survival rates and the
interaction cross section of the proposed planet, and \citet{Lawler2016}
investigated the impact of Planet Nine of the Kuiper belt objects.
\citet{Perets2012a} described a mechanism of a planet captured from
a different star system or a rouge planet, and predicted a wide, eccentric
and inclined orbit of the captured planet.

In this work we explore the interaction of a captured Planet Nine
with a debris disk in two scenarios: (a) an initially flat ecliptic disk around the Sun and (b) an initially inclined and eccentric disk. 
The disk we consider represents the early debris disk, remnant from the Sun's formation
era. For a short-term interaction of a planet with a ring of debris
around it, see \citet{Lee2016}. We show that Planet Nine, due to
its large sma, dominates the evolution of the outer disk between $1000-5000{\rm AU}$.
This results in a spheroidal structure at $1100{\rm AU}\lesssim a_{{\rm TAUS}}\lesssim1500{\rm AU}$
where TAUS stands for ``Thousand AU Sphere''. For scenario (a) this structure is surrounded by an inclined
disk aligned with Planet Nine's orbital plane and a warped disk towards
the ecliptic plane. For scenario (b) the TAUS and inclined disk are created but less visible due to the initial conditions. 
This structure serves as a new source of Sun grazing
comets, penetrating the inner solar system by long term secular evolution
and dynamical interaction with Planet Nine.

Our paper is organized as follows: in section~\ref{sec:Numerical-experiments}
we describe the numerical simulation. In section~\ref{sec:Results}
we describe the results of the numerical runs in comparison to the
standard Hills cloud formation mechanism, i.e. a fly-by event in the
early stages of the solar system. In section~\ref{sec:Implications-and-Discussion}
we present the implications of TAUS to the morphology of the outer
solar system as well as the rate and orientation of comets originating
from this region. Finally, we summarize the results and their implications
in section~\ref{sec:Summary}.

\section{Numerical experiments}
\label{sec:Numerical-experiments}

We used the code MERCURY6 \citep{Chambers1999}, N-body code with
a the Bulirsch-Stoer algorithm. The timestep used was $185\ {\rm days}$
which is one thousandth of the period of Planet Nine around the Sun
$\left(\sim10^{-3}P_{9}\right)$ and accuracy parameter of $10^{-12}$.
Only the Sun and Planet Nine were treated explicitly while the rest
of the four giant planets (Jupiter, Saturn, Uranus and Neptune) were
included secularly via their $J_{2}$ moment term 
\begin{equation}
J_{2}=\frac{1}{2}\sum_{i=1}^{4}\frac{m_{i}a_{i}^{2}}{MR^{2}}
\end{equation}
where $m_{i}$ and $a_{i}$ are the mass and its sma of planet number
$i$ respectively. $R=a_{4}$ is the effective radius of the inner
solar system which was set to be the orbital sma of Neptune. Any particle
that crossed $R$ (from outside) was removed from the integration.
We can neglect the Galactic tides effects up to the scale of $\sim10^{4}{\rm AU}$
\citep{Veras2013}.

For scenario (a) we simulated an initially flat and circular debris disk around the
Sun with $32,000$ massless particles of surface density profile $\sigma\propto a^{-\gamma}$,
with $\gamma=-1$ and $a$ being the massless particle sma \citep{Andrews2010}.
\citet{Larwood2000} and \citet{Kalas2000} use a different power-law
index, $\gamma=-3/2$; this value has only a minor effect on the results
presented in section \ref{sec:Results}. The region of the disk encompasses
$700{\rm AU}<a<7000{\rm AU}$, with eccentricity chosen from a flat
distribution in the range $e\in\left\{ 0,0.1\right\} $. The initial
inclination $i=0$, so that the longitude of the ascending nodes,
$\Omega$, and the argument of the periapsis, $\omega$, are ill defined.
The mean anomaly $\mathscr{M}$ was drawn from a flat distribution
of angles in the range $\mathscr{M}\in\left\{ 0,360^{\circ}\right\} $.
For senario (b) we simulated $32,000$ massless particles with surface density and sma distribution as in senario (a). 
We used thermal distribution for the eccentricity and inclination drawn uniformly from 
a distribution with aspect ratio of $0.1$, for the vertical distance over the distance from the Sun.
The argument of pericentre the longitude of the ascending node and the mean anomaly were drawn randomly from
a flat distribution of angles between $0$ and $360^{\circ}$.

Planet Nine parameters were taken from \citet{Batygin2016} and \citet{Brown2016},
with $m_{9}=10m_{\oplus}$, $a_{9}=700{\rm AU}$, $e_{9}=0.6$, $i_{9}=30^{\circ}$,
$\omega_{9}=130^{\circ}$, $\Omega_{9}=0$ and $\mathscr{M_{9}}=0$.
Since the giant planets are incorporated through $J_{2}$ and the
disk has cylindrical symmetry, the exact value of $\Omega_{9}$ has
no physical significance. 

We integrated the system for $4{\rm Gyr}$ and followed the change
of the orbital parameters as well as morphology of the debris disk.
The results from these simulations are presented in the next section.

\section{Results}
\label{sec:Results}

In this section we present the numerical integration results for the two scenarios described
in section \ref{sec:Numerical-experiments}. Subsection \ref{sub:Scenario_a} is dedicated for scenario (a) i.e. flat and circular 
initial disk. In subsection \ref{sub:Scenario_b} we present the results for scenario (b), namely for initially inclined and eccentric disk.
Both subsection present the results  after integrating the system for $4{\rm Gyr,}$ and in both we find three qualitatively different regions
of the disk: (\mbox{i}) The most inner stable part of the disk has
a \textit{spheroidal structure} (TAUS); (\mbox{ii}) beyond the sphere
there is an \textit{inclined disk} with respect to the ecliptic; and
(\mbox{iii}) beyond the inclined part of the disk, reside a disk warped
back to the ecliptic plane.

\subsection{Scenario (a)}
\label{sub:Scenario_a}

\begin{figure*}
\includegraphics[width=17cm]{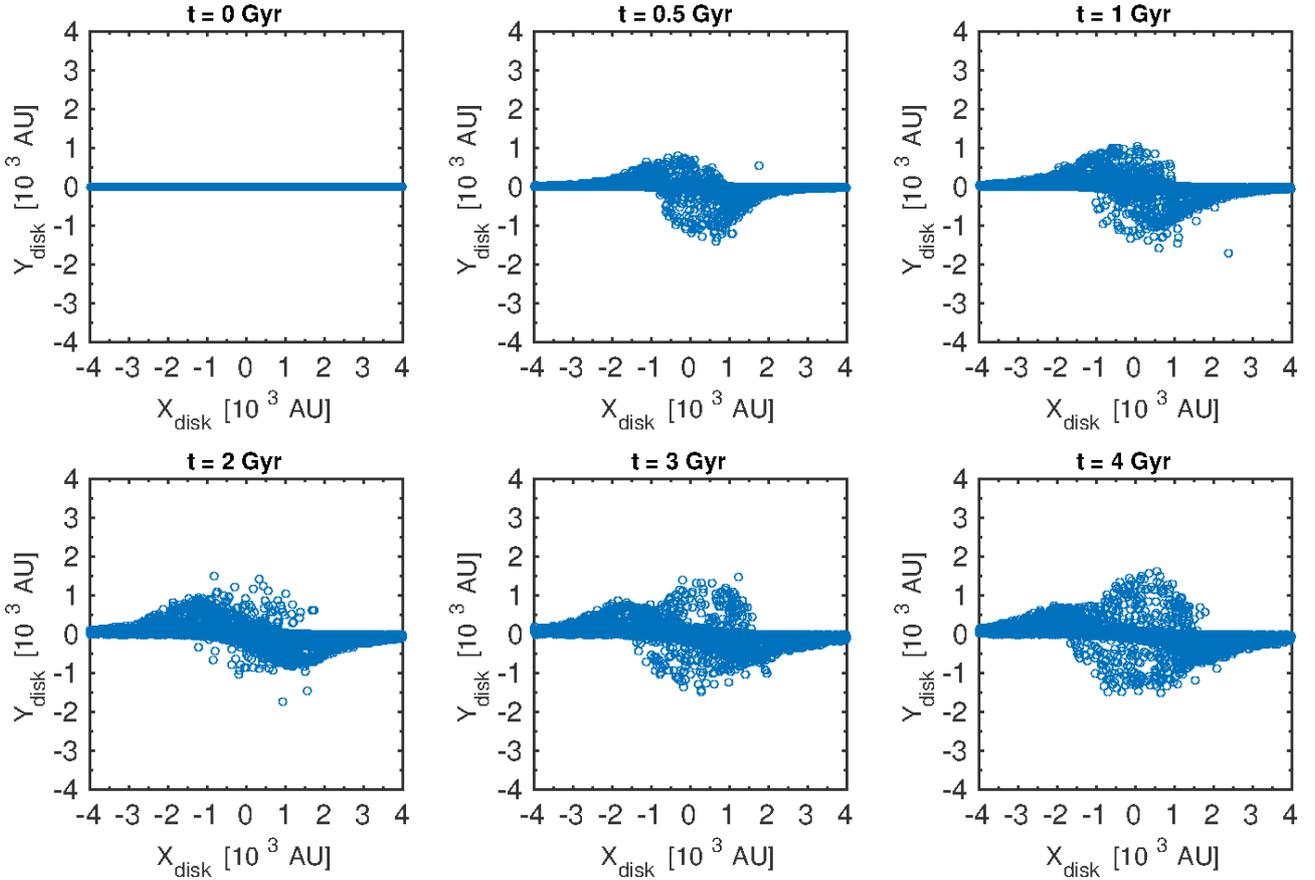}

\protect\caption{\label{fig:Edge-on}Edge on view of the outer solar system at $0.5,1,2,3,4{\rm \ Gyr}$
after formation. The centre notes the Sun and the blue circles are
the massless particles. Panel (a) provides a snapshot of the initial
conditions for a flat non inclined disk with inner cutoff of $700{\rm AU}$
and an outer cutoff of $7000{\rm AU}$. At the final snapshot $t=4{\rm Gyr}$,
all three regions of the disk are visible: (\mbox{i}) the TAUS at
a radius of $\sim1500{\rm AU}$; (\mbox{ii}) an inclined disk, about
$20^{\circ}$ from the ecliptic plane, between $1500{\rm AU}<a<3000{\rm AU}$
and (\mbox{iii}) a warped disk relative to the ecliptic plane, at
$a\gtrsim3000{\rm AU}$. }
\end{figure*}

\begin{figure}

\includegraphics[width=8cm]{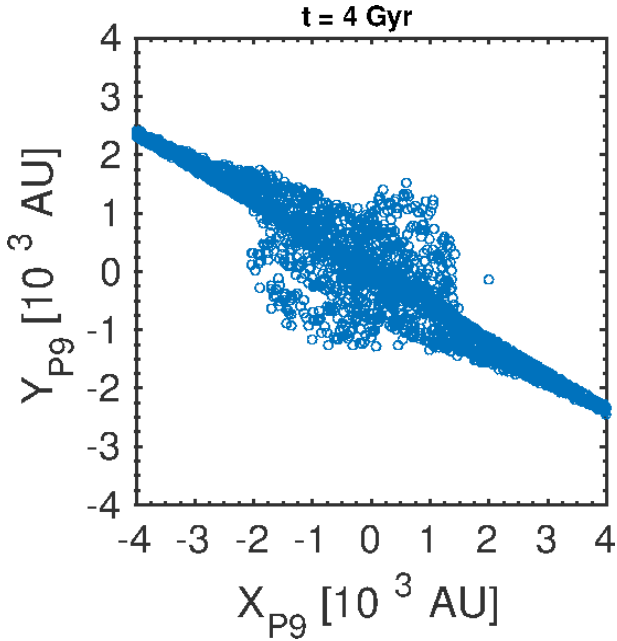}\protect\caption{\label{fig:Aligned2PN}The same as Figure \ref{fig:Edge-on} with
the y-axis aligned with Planet Nine's orbital angular momentum. The
line of node is perpendicular to the displayed plane.}

\end{figure}

Figure \ref{fig:Edge-on} presents several snapshots from the scenario (a)  at an edge-on view relative
to the ecliptic plane in the outer solar system at different times,
with the x-direction chosen arbitrarily and the y-direction is aligned
with the orbital angular momentum of the ecliptic plane. Figure \ref{fig:Aligned2PN}
presents an edge-on view from Planet Nine's orbital plane. At this
orientation, Planet Nine's orbital angular momentum is pointing in
the y-direction and the line of nodes is perpendicular to the origin.
The spheroidal structure is visible from both perspectives.

\begin{figure*}
\includegraphics[width=8cm]{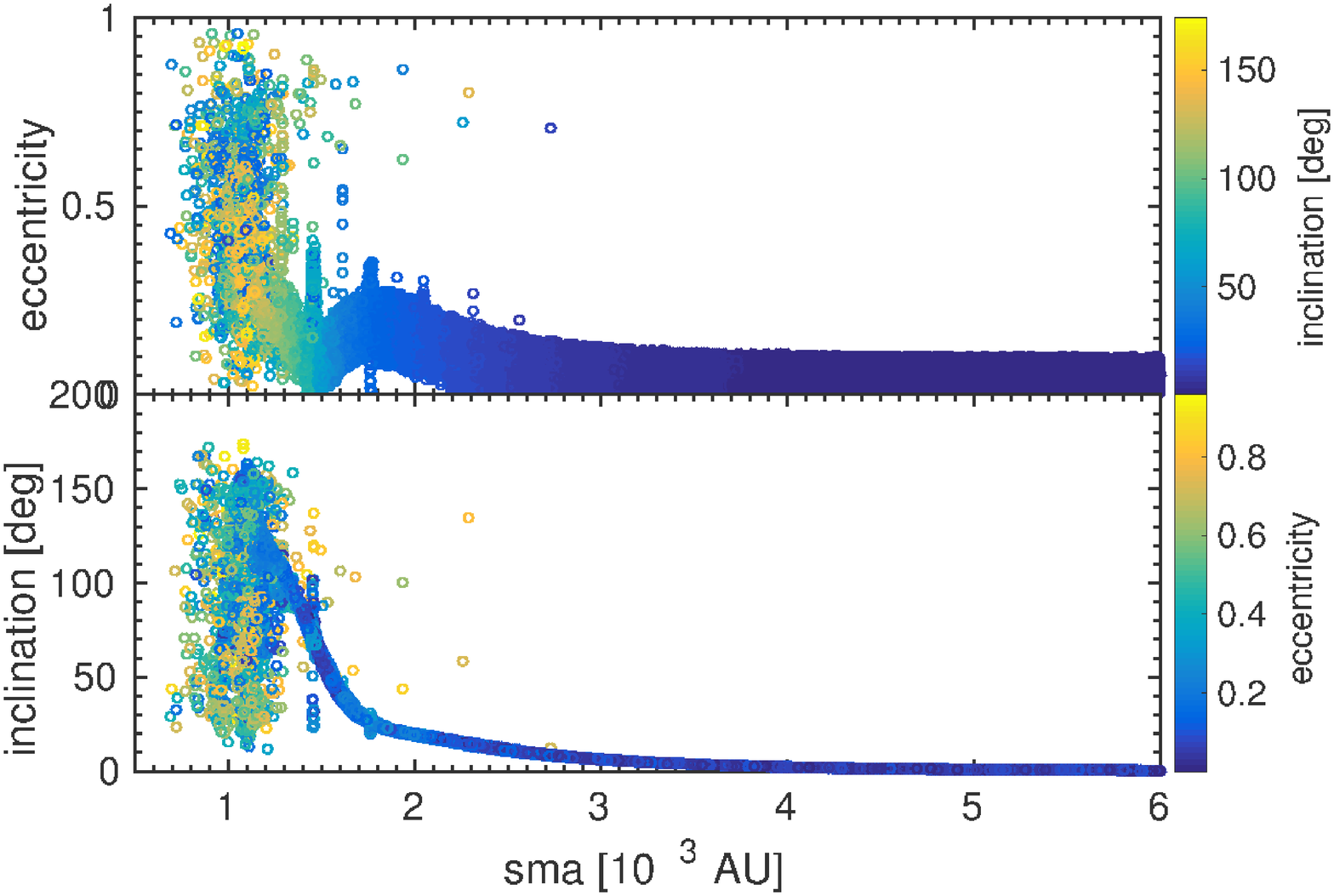}$\hfill$\includegraphics[width=8cm]{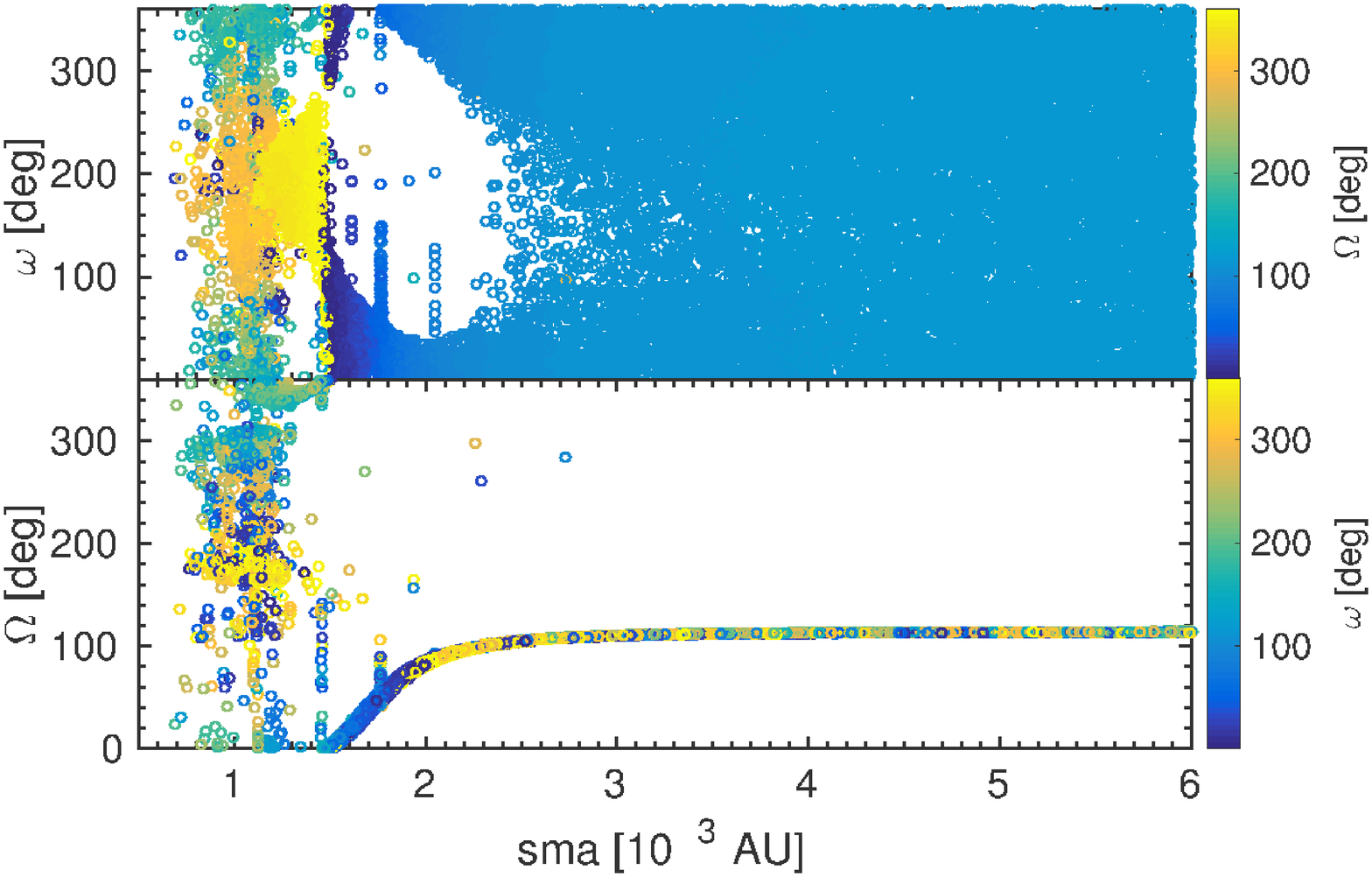}
\protect\caption{\label{fig:inc_ecc_omega_Omega_circ}\textbf{Left panel}: the final inclination and eccentricity
distributions as a function of sma, with eccentricity and inclination, respectively, colour coded.
\textbf{Right panel:} the final argument of pericentre and the longitude of the ascending node final distribution as a function of sma. }
\end{figure*}

\begin{figure*}
\includegraphics[width=16cm]{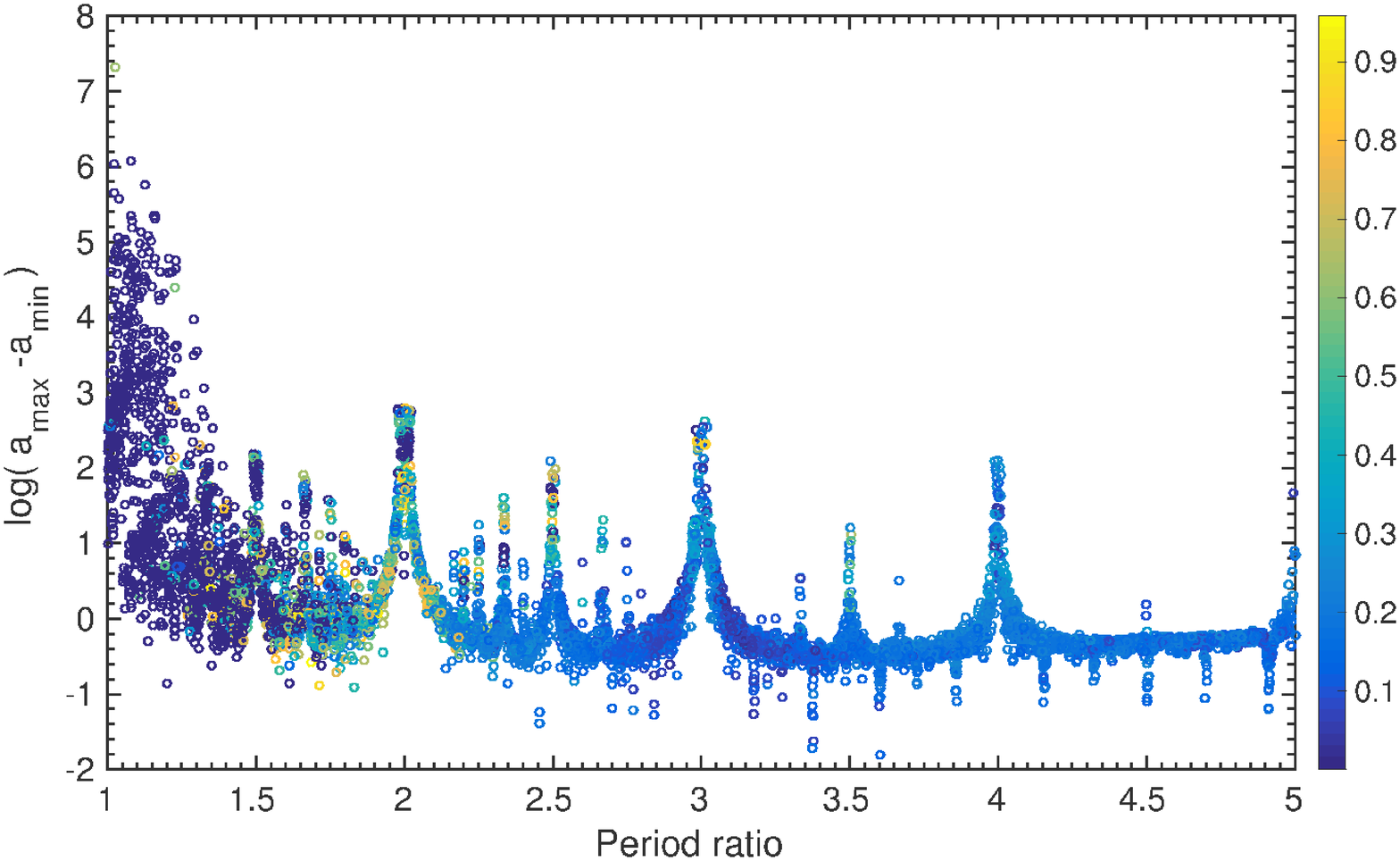}\protect\caption{\label{fig:circ_log_delta_a}  A measure of the maximal change in sma as a function 
of period ratio. The eccentricity is colour coded. }
\end{figure*}

The structure of the disk is complex. Figure \ref{fig:inc_ecc_omega_Omega_circ} presents the eccentricity, inclination, argument of pericentre and the 
longitude of the ascending node of the test particles as a function of the sma. The results exhibit a spread in inclination, eccentricity and the longitude 
the ascending nodes. This implies a spherical like structure up to  $1500 \rm{AU}$ which decline gradually with distance.
The complexity originates from mean motion resonances (MMRs) with Planet Nine and secular interaction
with Planet Nine's orbit \citep{Murray1999}. 
MMRs are caused when the ratio of the orbital periods $P_{1}$
and $P_{2}$ is a rational number, 
\begin{equation}
\frac{\alpha}{\beta}=\frac{P_{1}}{P_{2}}=\left(\frac{a_{1}}{a_{2}}\right)^{3/2},
\end{equation}
with $\alpha$ and $\beta$ being (small) integers. In MMRs the resonant angle $\phi$ defined as 
\begin{equation}
\phi=j_1\lambda_9+j_2\lambda+j_3\varpi_9+j_4\varpi+j_5\Omega_9+j_6\Omega,
\end{equation}
where $\lambda$ is the mean longitude, $\varpi$ is the longitude of pericentre and
$\Omega$ is the longitude of the ascending node. The prefactors $j_i$ satisfies 
\begin{equation}
\sum_i j_i=0.
\end{equation}
A fingerprint of MMR involves the excitation of
orbital parameters at specific period ratio with the basic frequency, i.e. the mean motion frequency of Planet Nine. 
The vertical lines in figure \ref{fig:e_i_omega_Omega_res_circ} are the locations of main MMRs with Planet Nine.
Different colour of vertical lines correspond to different order of resonance. The proximity of the resonances below $3:1$ suggests
that the dynamics in this regions are influenced by MMR overlap \citep{Morbidelli1995}. 
 The resonances above $3:1$ are isolat. Further evidence for the importance of the MMRs is presented in figure \ref{fig:circ_log_delta_a}.
 The distinction between isolated and overlapping MMRs is clear below the $3:1$ resonance.
 The analytic treatment of the MMRs goes beyond the scope of this manuscript.
The inner part TAUS shows clustering in $\omega$ and $\Omega$. The rest of the disk has two regions: the inclined region $1500{\rm AU}<a<3000{\rm AU}$ 
which is aligned with Planet Nine's orbital plane similar with  \citep{Mouillet1997} and the flat disk on the ecliptic plane.

\begin{figure*}
\includegraphics[width=8cm]{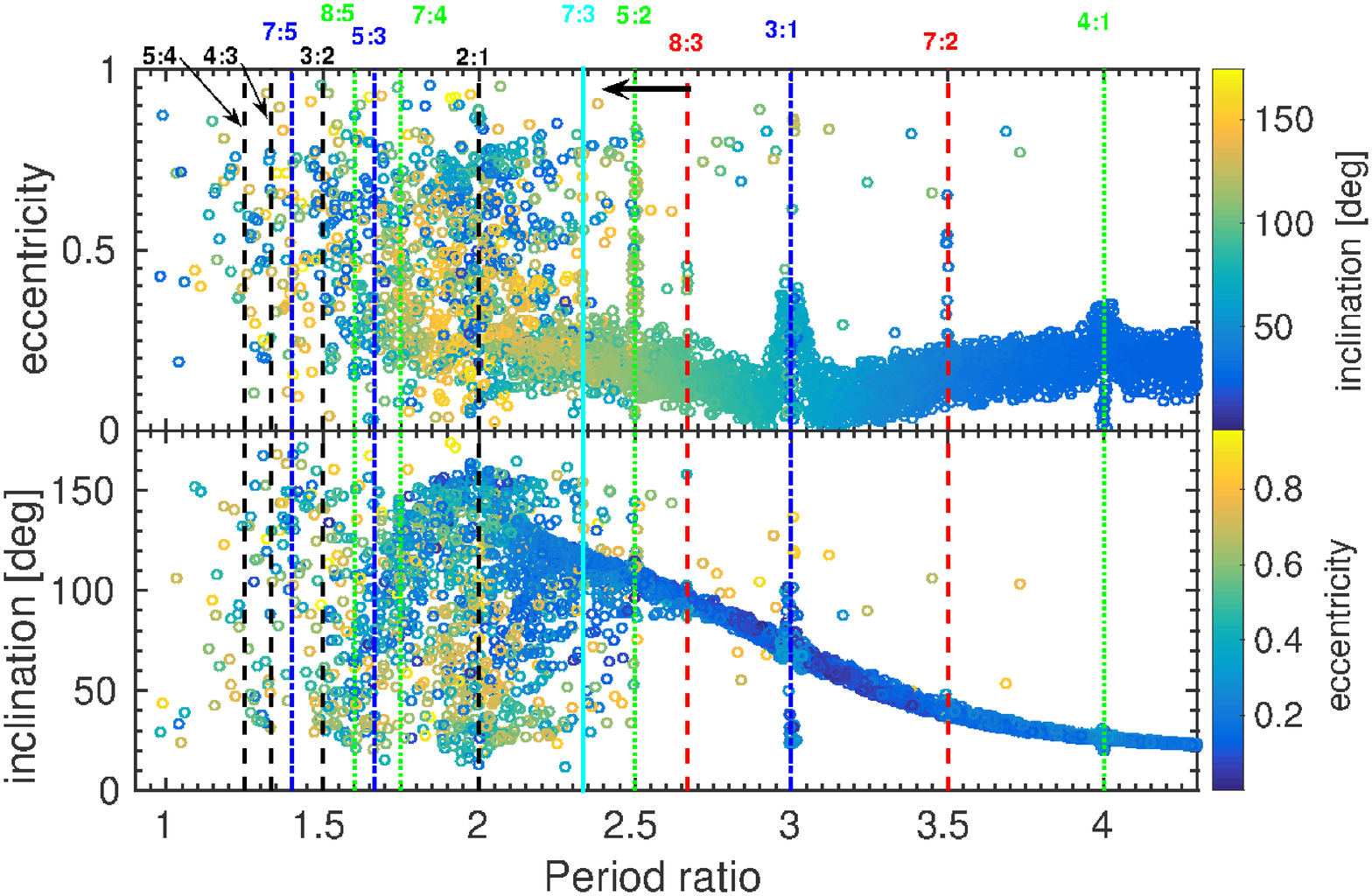}$\hfill$\includegraphics[width=8cm]{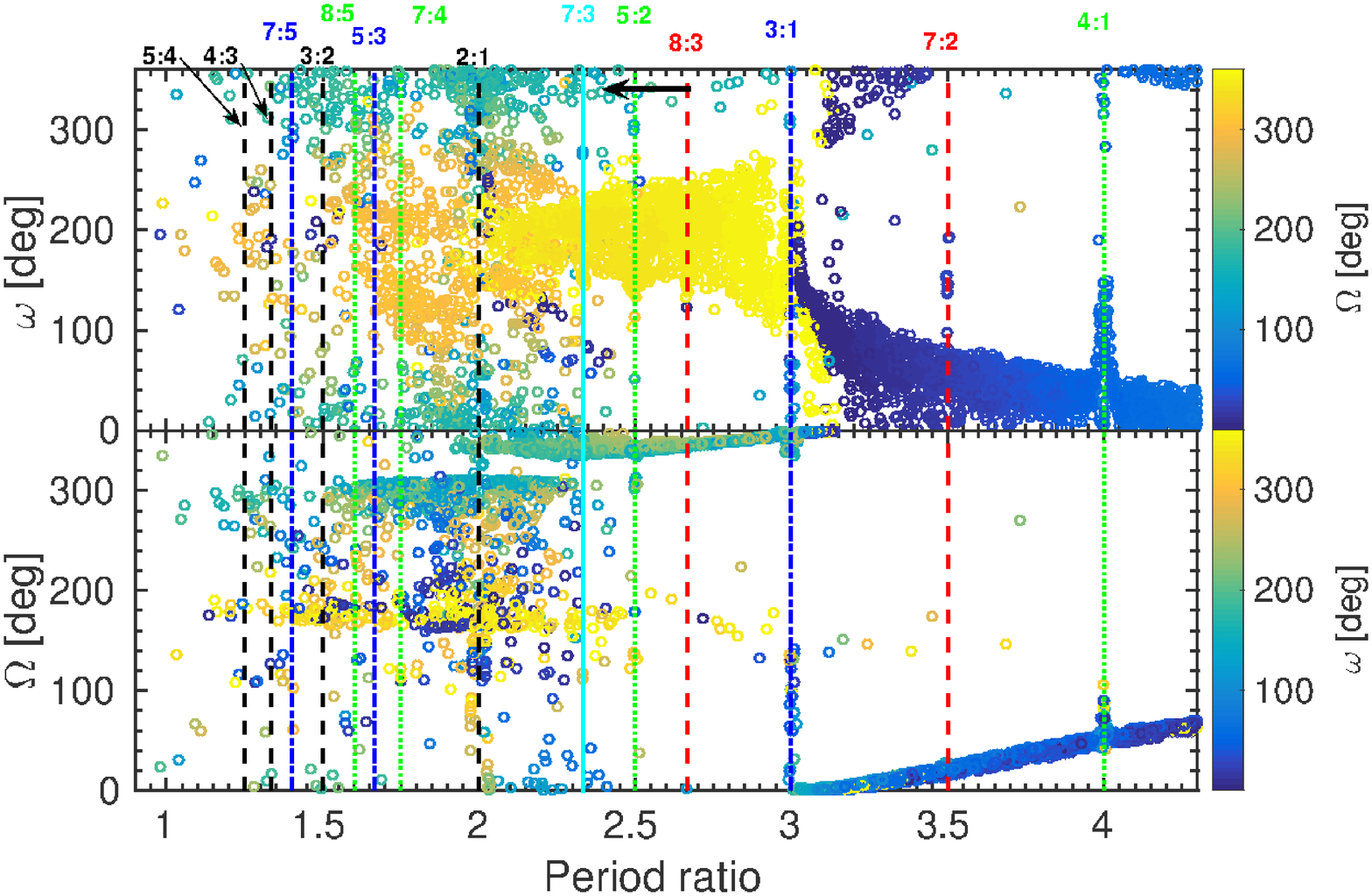}\protect\caption{\label{fig:e_i_omega_Omega_res_circ} \textbf{Left panel:} the final state
of eccentricities and inclinations as a function of period ratio. The vertical lines corresponds to the main MMRs. Order $1$ black; order $2$ blue; order $3$ green; order $4$ cyan; order $5$ red.
The black arrow indicate the end of the TAUS at $~1500 \rm{AU}$  \textbf{Right panel:}
 the final distribution of the argument of pericentre and the longitude of the ascending nodes. In both panel the MMRs are visible.}
\end{figure*}

\subsection{Scenario (b)}
\label{sub:Scenario_b}

\begin{figure*}
\includegraphics[width=8cm]{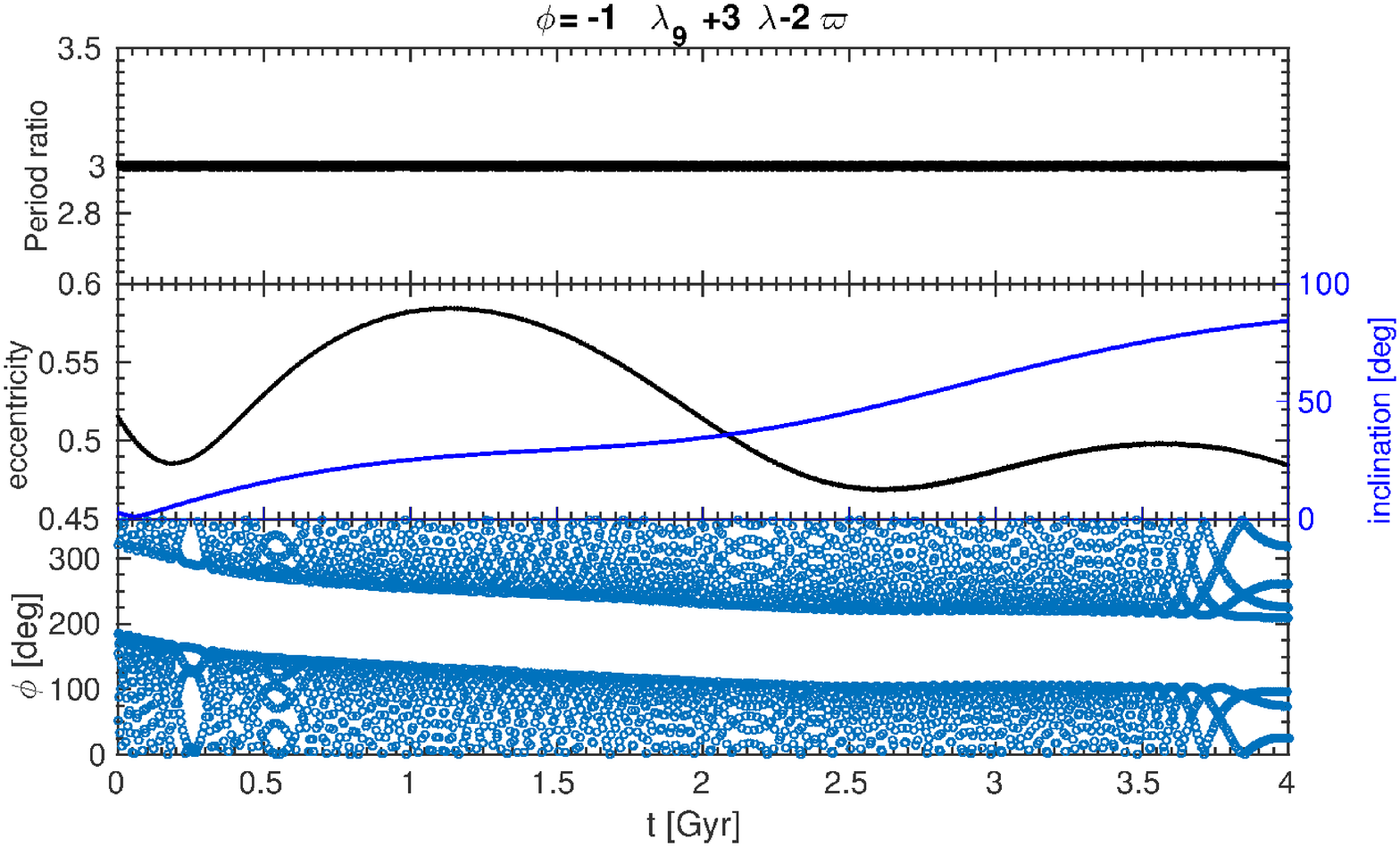}$\hfill$\includegraphics[width=8cm]{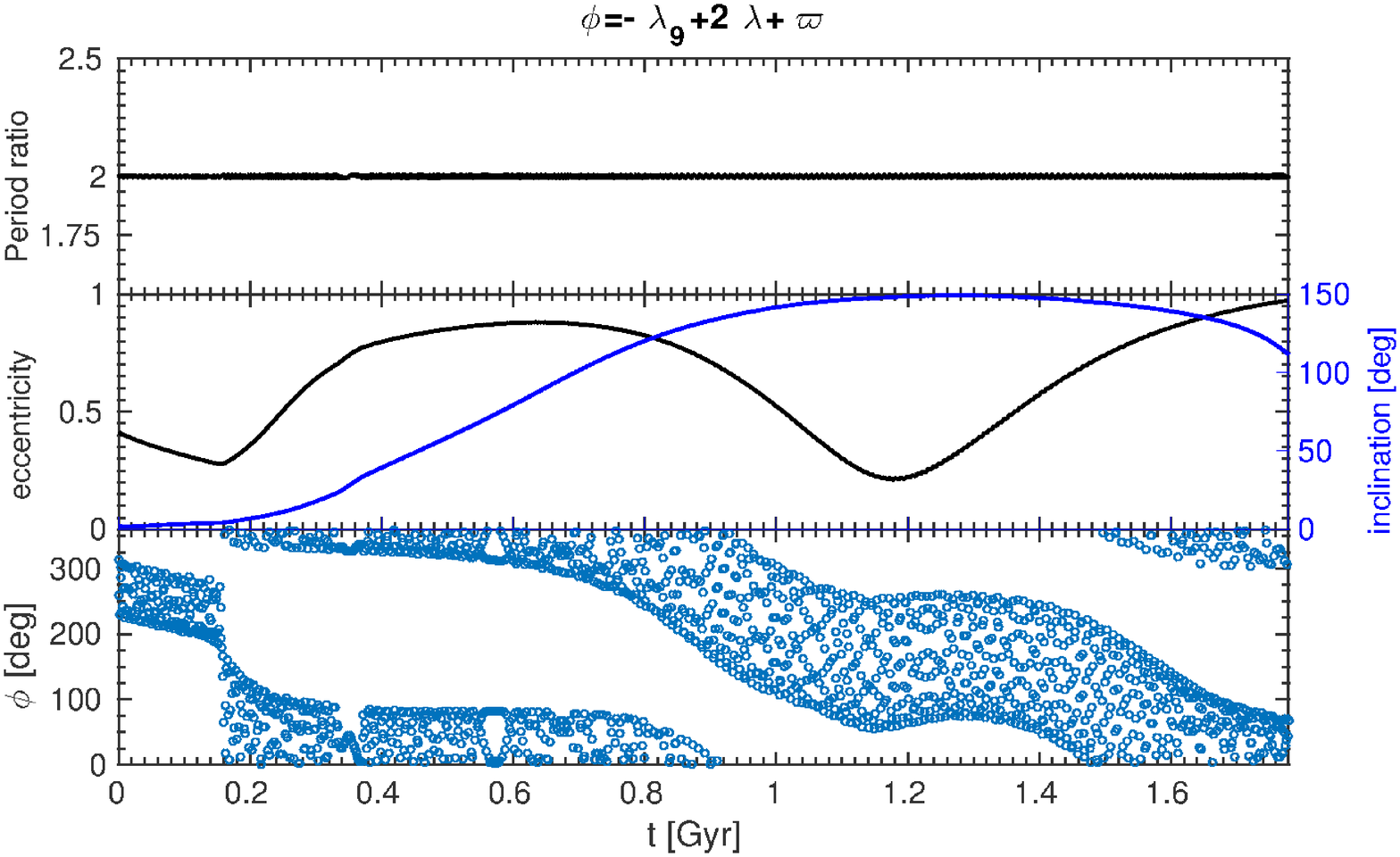}\protect\caption{\label{fig:examples_inclined} \textbf{Left panel:} The evolution
of single test particle. The particle is in a $3:1$ MMRs with the resonance angle of $\phi=-\lambda_9 +3\lambda-2\varpi$. The top panel is the period ratio as
a function of time. The middle panel in the eccentricity evolution (black left axis) and the inclination evolution (blue right axis). The bottom panel is the evolution
of the resonance angle $\phi$. \textbf{Right panel:} The evolution of an ejected particle at $2:1$ MMR with $\phi=-\lambda_9 +2\lambda-\varpi$. }
\end{figure*}

Next we present the structure of the disk in the second scenario, of an inclined and eccentric initial disk.
The underlining structure of the final state is similar to scenario (a). The basic structure of spherical TAUS and inclined disk 
is supplemented by hot and inclined group of test particles.
Figure \ref{fig:e_i_omega_Omega_res_circ} present the final orbital elements as a function of sma. Unsurprisingly the TAUS
is extended toward ~$3000 \rm{AU}$  distance. The cold inclined disk is visible in the inclination plot. Two specific sets of test particles 
are presented in Fig. \ref{fig:examples_inclined}. The left panel presents the evolution of an inclined particle while the right panel shows the evolution of ejected particles.
In both cases the resonant angle $\phi$ is plotted versus time.

\begin{figure*}
\includegraphics[width=8cm]{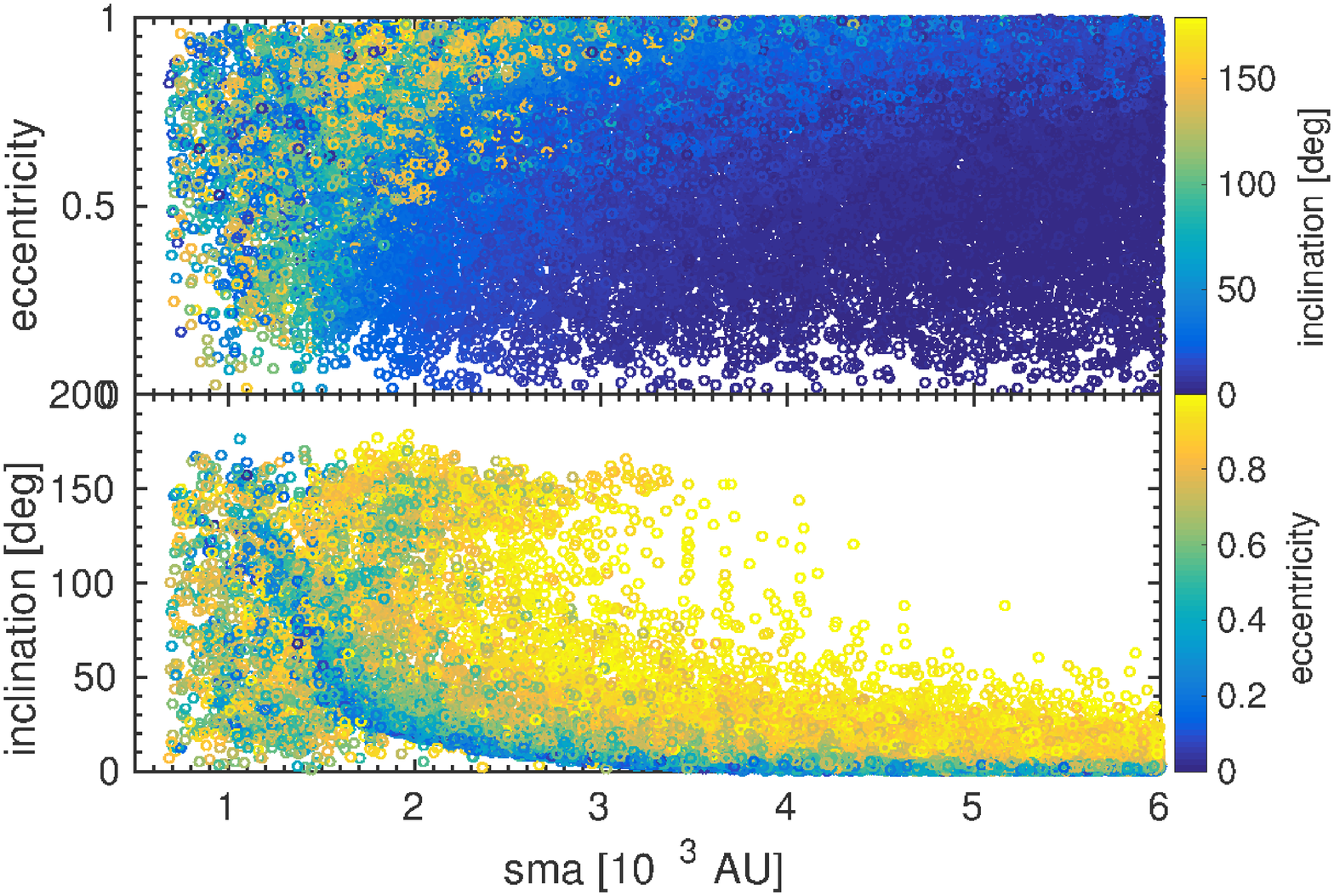}$\hfill$\includegraphics[width=8cm]{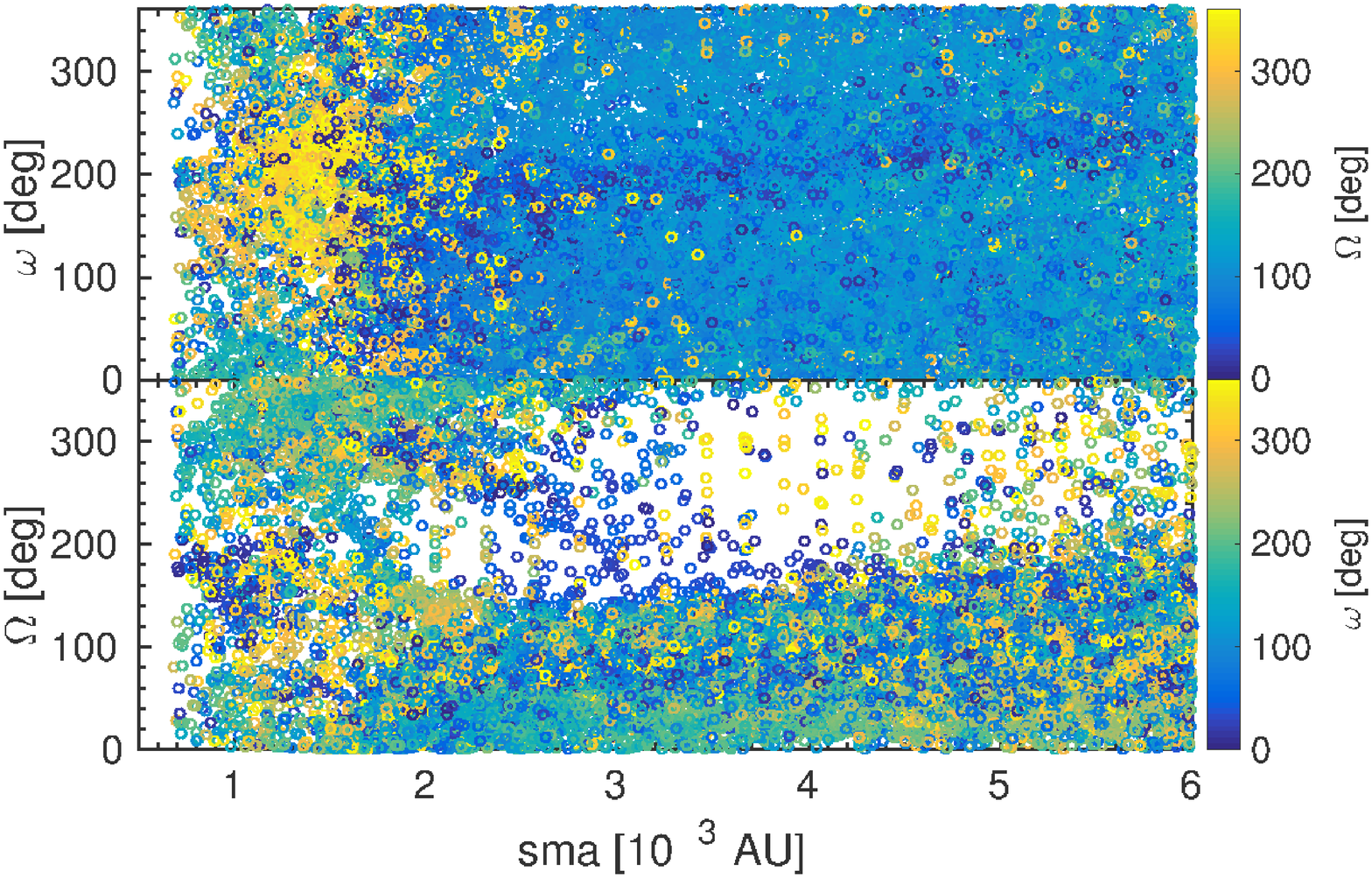}\protect\caption{\label{fig:e_i_omega_Omega_res_circ} \textbf{Left panel:} the final state
of eccentricities and inclinations as a function of period ratio for scenario (b). \textbf{Right panel:}
 the final distribution of the arguments of pericentre and the longitude of the ascending nodes.}
\end{figure*}

\begin{figure*}
\includegraphics[width=16cm]{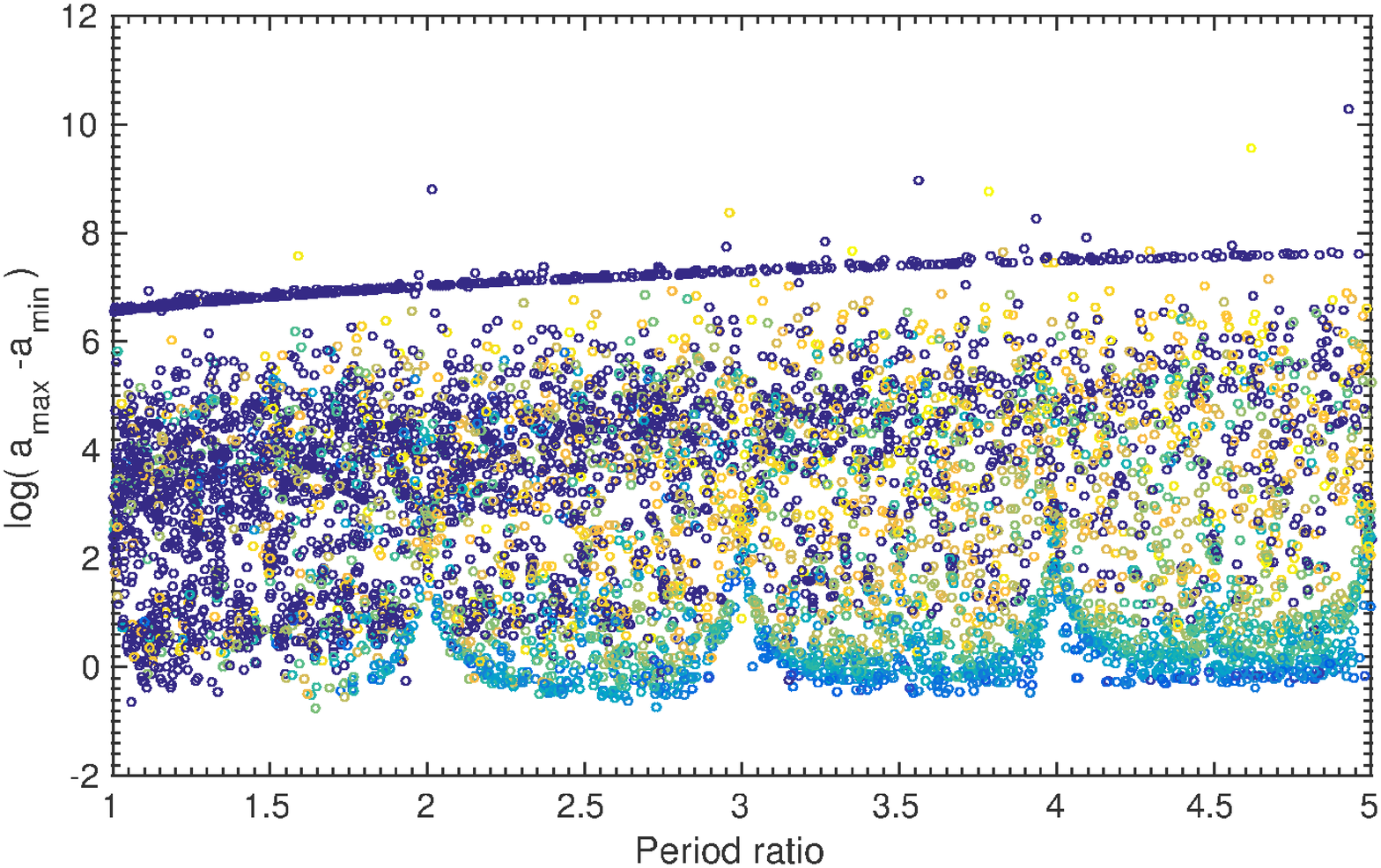}\protect\caption{\label{fig:inclined_log_delta_a}  A measure of the maximal change in sma as a function 
of period ratio. The eccentricity is colour coded. }
\end{figure*}

We also examine the final distributions in the properties of the
relevant orbital elements of the different disk regions (shown in both scenarios): TAUS, inclined
disk, and the warped+ecliptic disk. We consider each region separately.

\subsection{TAUS}
\label{sub:TAUS}
TAUS is the spheroidal region of the disk located between $1000{\rm AU}\apprle a\apprle1500{\rm AU}$.
The distribution of the eccentricities, $e$, the longitude of the
ascending node, $\Omega$, and the argument of periapsis, $\omega$,
are presented in the appendix at  \ref{fig:TAUS_ecc_inc_Omega_omega}. TAUS
maintains highly eccentric orbits, with $e>0.8$. Since particles
with periapsis distance of $q\leq a_{{\rm Neptune}}$ are removed
from the integration, the eccentricity cannot get arbitrary close
to unity and are bound by $e_{c}=1-a_{{\rm Neptune}}/a$, with $e_{c}\approx0.98$.

The longitude of the ascending nodes show two features. The first
is a narrow distribution for the majority of the test particles that
follows the precession induced by Planet Nine. Planet Nine's orbit
dominates the dynamical evolution of TAUS. The second is a wider distribution
of the minority of the particles as a result of the resonant interaction
with Planet Nine, as illustrated in Figure \ref{fig:TAUS_ecc_inc_Omega_omega}. 

The argument of periapsis has a wide distribution around $\sim200^{\circ}.$
The inclination distribution is also wide due to strong secular and
resonance interactions with Planet Nine. In the TAUS, there are prograde
and retrogrades orbits. This wide distribution of inclinations give
TAUS its spheroidal shape.

\subsection{Inclined disk}
\label{sub:Inclined-disk}
The inclined disk is relatively flat and has an inclination with a mean value $\left\langle i\right\rangle \approx18.4^{\circ}$
with respect to the ecliptic. It is located in a region between the
TAUS and the ecliptic disk, namely $1500{\rm AU}\lesssim a\lesssim3000{\rm AU}$.
The secular interaction with Planet Nine torques the disk and aligns
it towards the orbital plane of Planet Nine \citep{Mouillet1997}. The alignment timescale
is of the order of $\sim{\rm Gyr},$ as evident in Figure \ref{fig:Edge-on}.
Figure \ref{fig:InclinedDisk-Omega-omega-inc} in the appendix presents the orbital
distribution of the test particles that are bounded between $1500{\rm AU}\lesssim a\lesssim3000{\rm AU}$.
The inclined disk has a preferred orientation in space; the argument
of periapsis is clustered around a value of zero (see Figure \ref{fig:InclinedDisk-Omega-omega-inc})
and the longitude of the ascending node is peaked around $\sim100^{\circ}.$

\subsection{Outer disk}
\label{sub:Outer-disk}
The outer disk includes test particles that did not interact with
Planet Nine at $a>3000{\rm AU}$. The particles show no sign of a
change in their initial orbital elements distribution. In other words,
the eccentricity, inclination, argument of periapsis and longitude
of the ascending nodes distributions, are the same as the initial
distributions, as shown in the appendix at Figure \ref{fig:Outer disk hist}.

\subsection{Comparison to a stellar encounter}
\label{sub:Comparison-to-a}
Next, we compare the signature of the Planet Nine - disk interaction
with the perturbation from a close fly-by of a star. We focus on plausible
stellar encounters in the local stellar density in the field, $n_{*}=0.1{\rm pc^{-3}}$
and a velocity dispersion of stars of $\sigma_{*}=40{\rm kms^{-1}}$
\citep{Binney2008} with a perturber mass of $m_{p}\approx0.5m_{\odot}$.
We note that in the early star cluster phase that lasts about $100{\rm Myr}$,
the conditions could have been completely different with three orders
of magnitude higher density $n_{*}\approx100{\rm pc^{-3}}$ and a
much smaller velocity dispersion of $\sigma_{*}=1{\rm kms^{-1}}$
\citep{Li2016}. We focus just on the stellar encounter in the field
because the solar system planets were still migrating during the early
cluster phase according to popular Nice model \citep{Levison2008}.

The time between stellar encounters, $t_{{\rm enc}}$, is the inverse
of the rate, which in turn proportional to the stellar density, $n_{*},$
the geometrical cross section, $\sigma_{{\rm cross}}=\pi b^{2}$,
where \textbf{$b$} is the impact parameter, and the relative speed
at infinity, $v$ 
\begin{equation}
\Gamma=n_{*}\sigma_{{\rm cross}}v=n_{*}\pi b^{2}v.\label{eq:encounter_rates}
\end{equation}
Gravitational focusing add a correction to the effective impact parameter
\begin{equation}
b_{{\rm focus}}^{2}=b^{2}\left[1+\left(\frac{v_{{\rm esc}}}{v}\right)^{2}\right],
\end{equation}
where $v_{{\rm esc}}$ is the escape velocity at distance $b$. For
$b=1000{\rm AU}$, $v_{{\rm esc}}\approx1.3{\rm kms^{-1}}$, yielding
negligible gravitational focusing.

Next we calculate $t_{{\rm enc}}=1/\Gamma.$ For the relevant impact
parameter of \textbf{$b\approx1000{\rm AU}$ }\citep{Hills1981},
the time between encounters is $t_{{\rm enc}}\sim3.3{\rm Gyr}$, namely
roughly once during the lifetime of the solar system. 

\begin{figure*}
\includegraphics[width=17cm]{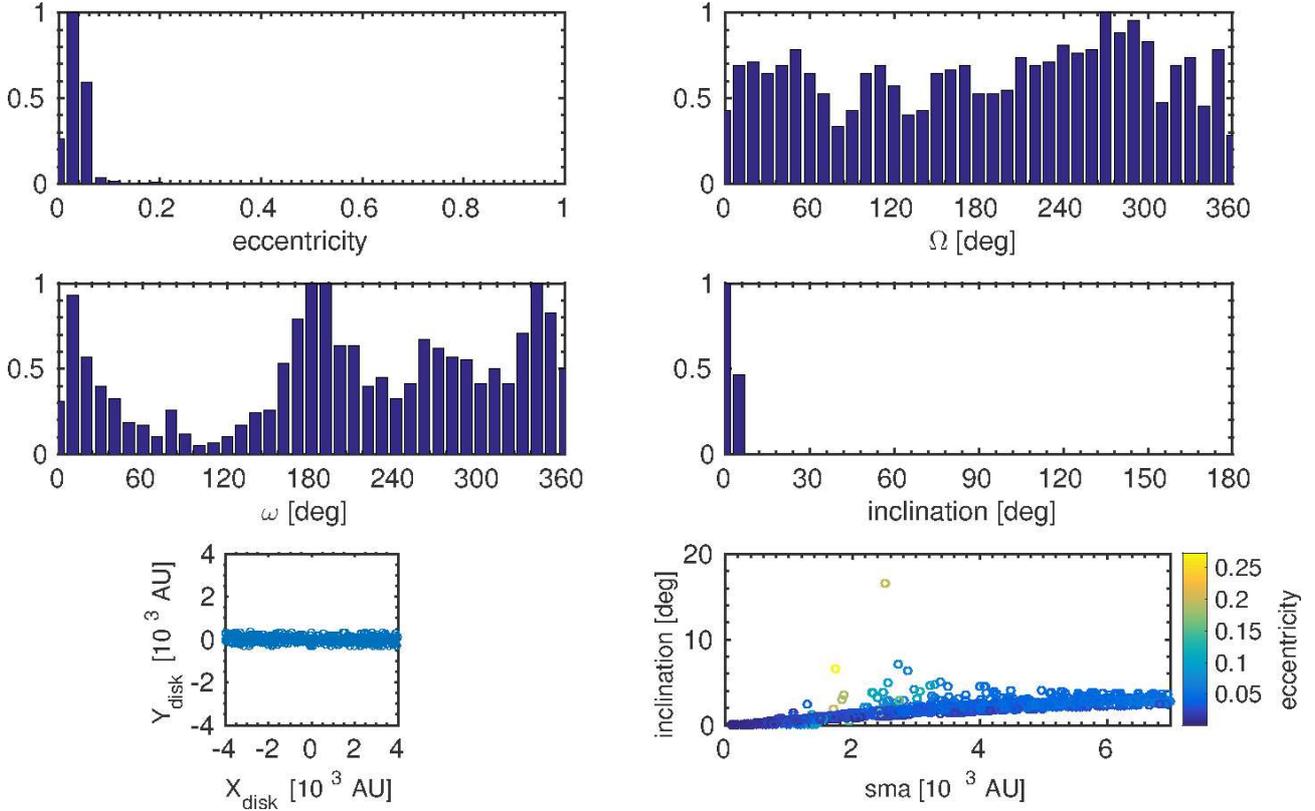}

\protect\caption{\label{fig:FlyBy-simulations}\textbf{Upper left:} The eccentricity
distribution is sharply around $\sim0.05.$ \textbf{Upper right:}
The longitude of ascending nodes, $\Omega,$ has a near uniform distribution.
\textbf{Middle left:} The argument of periapsis, $\omega,$ has a
wide distribution. \textbf{Middle right:} The inclination distribution.
The disk remains flat.\textbf{ Bottom left:} Final snapshot of the
disk after a stellar encounter. The disk is on the X-Z plane. \textbf{Bottom
right:} The inclination as a function of sma (the eccentricity colour
coded) shows a weak dependence on sma.}

\end{figure*}

We simulate such an encounter using MERCURY6 \citep{Chambers1999}
with the Sun and the four giant planets contribution through $J_{2}$.
The disk around the Sun is modeled by $1000$ massless particles in
the same way described in section \ref{sec:Numerical-experiments}.
Additionally, we modeled the stellar encounter to be with $m_{p}=0.5m_{\odot}$
ans a hyperbolic trajectory with periapsis distance of $q_{p}=1000{\rm AU}$,
orbital inclination $i_{p}=90^{\circ}$, and an argument of periapsis,
$\omega_{p}=60^{\circ}$. We gave the perturbing star the trajectory
that corresponds to $v_{\infty}=40{\rm kms^{-1}}$ and integrated
the system for $1{\rm Myr}$, the orbital properties and the disk
morphology are stable after the stellar encounter without any other
strong dynamical interactions. Figure \ref{fig:FlyBy-simulations}
shows our simulation results. Due to the high relative velocity the
interaction is extremely weak. The final snapshot from an edge-on
projection reveals the main qualitative difference from the Planet
Nine scenario. The lack of spheroidal structure is clear visible. 

\subsection{Ejected particles}
\label{sub:Ejected-particles}
\begin{figure*}
\includegraphics[width=17cm]{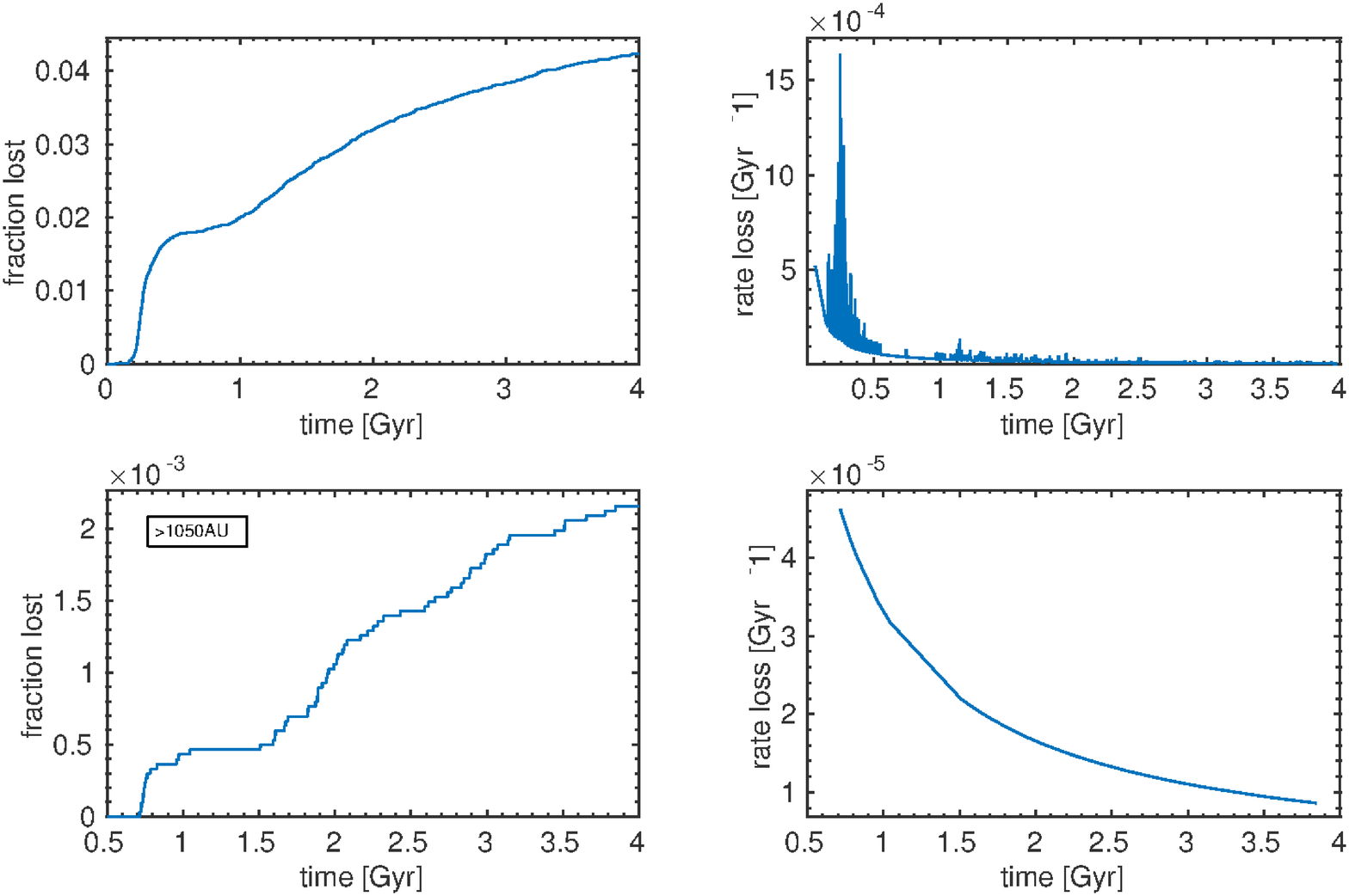}

\protect\caption{\label{fig:Loss Fraction}\textbf{Upper left:} Loss fraction from
the entire population of particles. During the integration time $\sim2\%$
of the particles were removed after entering the inner solar system.
\textbf{Upper right:} The rate of ejection as a function of time.
This panel is the time derivative of the left panel. \textbf{Bottom
left:} Loss fraction from the overall particles with initial sma greater
than $1050{\rm AU}$. \textbf{Bottom right:} The rate of ejection
as a function of time. Both plots shows a high loss rate at $\sim300\ {\rm Myr}$
which implies large number of objects interacting with Planet Nine
in this era.}
\end{figure*}

During our simulation, several particles are dynamically unstable
and reach sufficiently high eccentricities $e>e_{c}$ to be removed
from our simulation. In Figure \ref{fig:Loss Fraction} we present
the lost fraction of particles in two representations. First, we show
all $32,000$ particles and the corresponding loss rate, i.e. the
time derivative of the lost fraction. Second, we show the fraction
of all particles that initially have $a>1050{\rm AU}$ as a function
of time and the rate of these events per year. 

\begin{figure*}
\includegraphics[width=8cm]{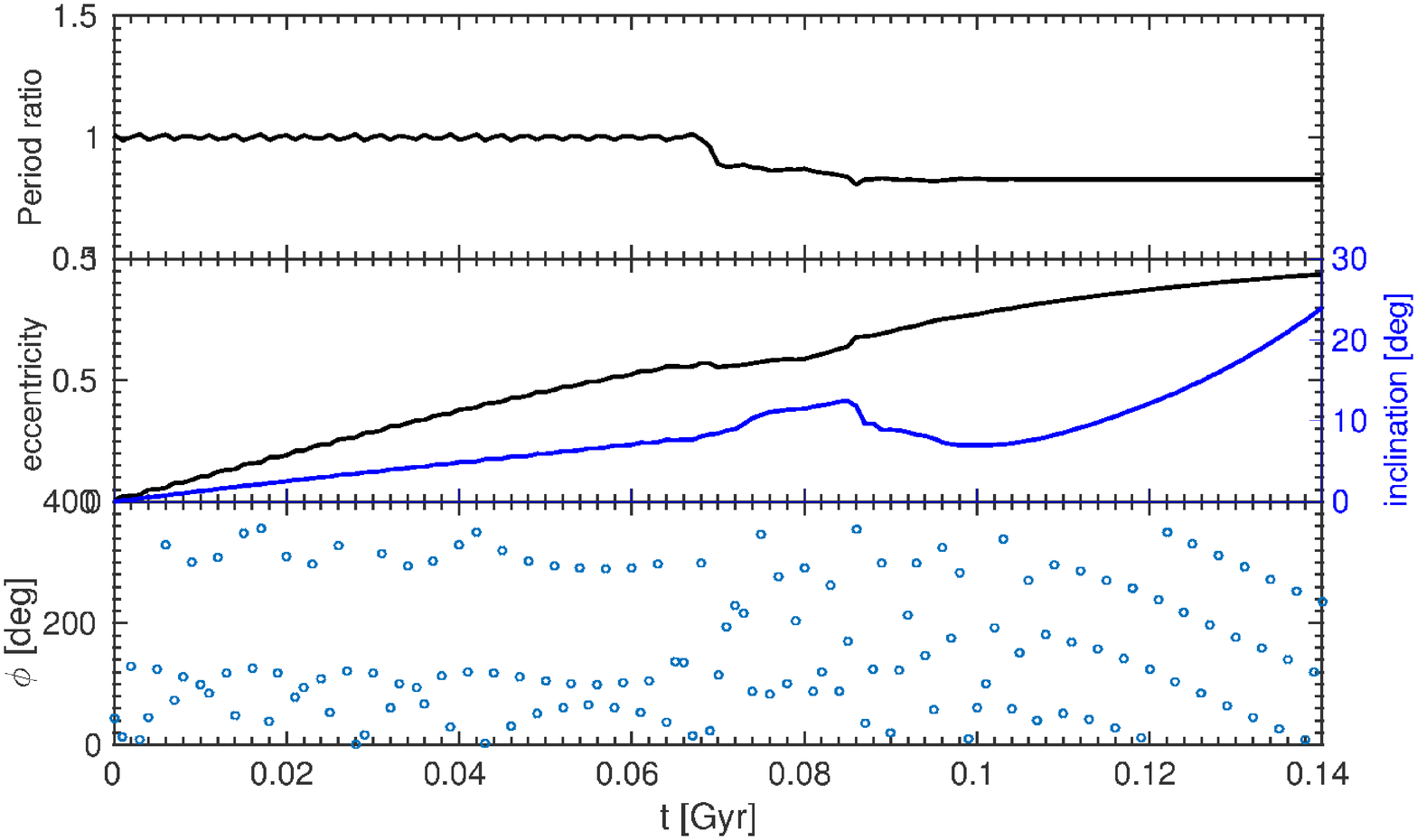}$\hfill$\includegraphics[width=8cm]{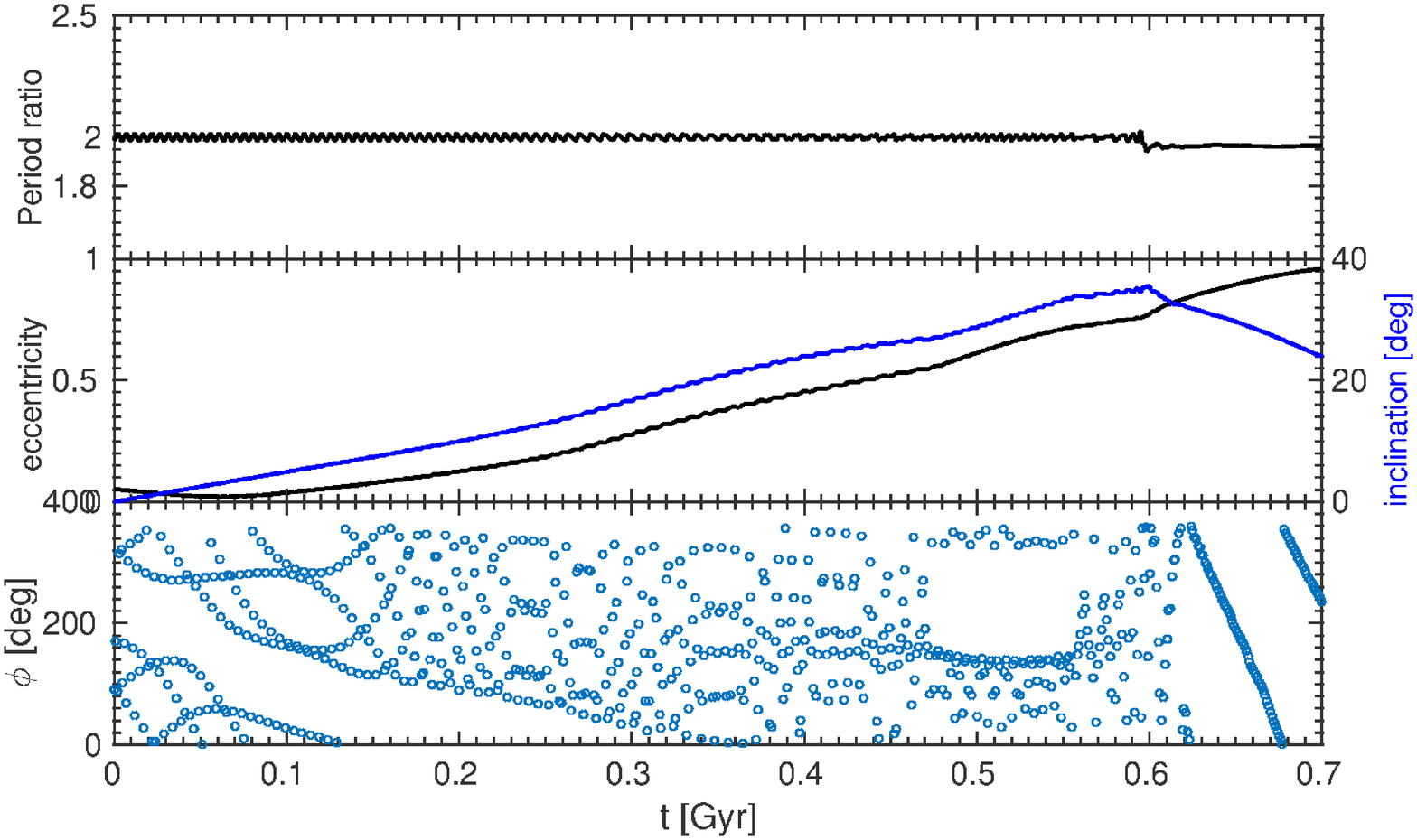}\protect\caption{\label{fig:ejection examples} The same as
figure \ref{fig:examples_inclined}. \textbf{Left panel:}  The particle is in a $1:1$ MMRs with Planet Nine.  \textbf{Right panel:}
The evolution of an ejected particle at $2:1$ MMR with $\phi=-\lambda_9 +2\lambda$.}
\end{figure*}

Figure \ref{fig:ejection examples} presents the evolution of two particles
that were removed from the system due to their high eccentricity. These particles were ejected
at different times from the simulation and become long period comets.
These objects originate from the TAUS, and represent a new population of comets. The sma of
the removed particles is $a\sim1000{\rm AU}$ and $700{\rm AU}$, respectively.

\begin{figure*}
\includegraphics[width=17cm]{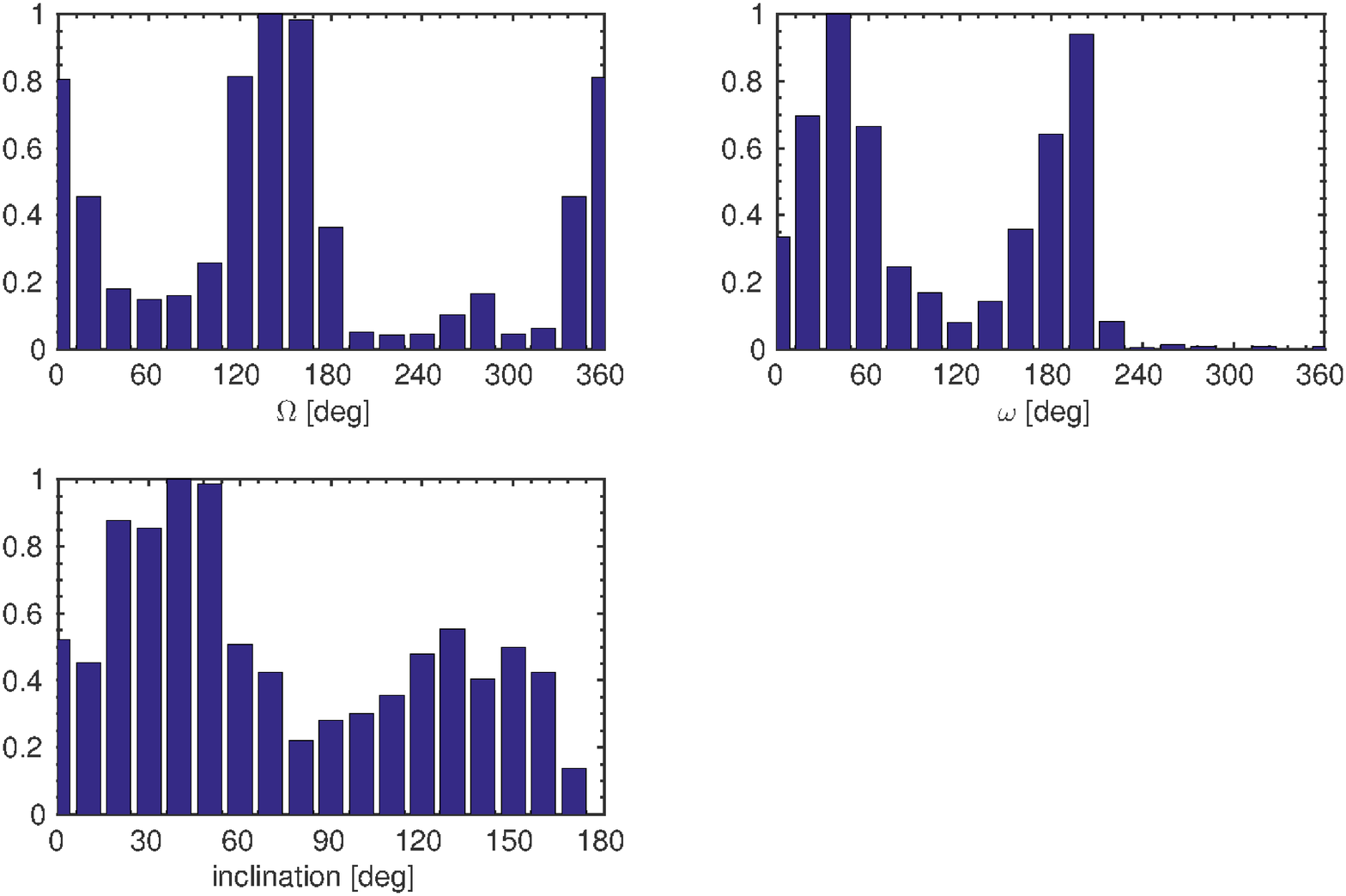}\protect\caption{\label{fig:Orbital-parameter-distribution-ejected}Orbital parameter
distributions of the removed particles from the simulation. The argument
of periapsis has two sharp peaks around: $\sim40^{\circ}$ and $\sim140^{\circ}.$
The inclination distribution is wide with two wide bumps around: $\sim20^{\circ}-50^{\circ}$
and $\sim120^{\circ}-150^{\circ}.$ The longitude of ascending nodes
has two sharp bumps around: $\sim0$ and $\sim140^{\circ}$. All distribution
are normalized a maximum value of unity.}

\end{figure*}

From our simulation we can infer the production rate of comets enter
the inner solar system. We present the distribution of orbital parameters
for all particles ejected from the simulation in Figure \ref{fig:Orbital-parameter-distribution-ejected}.
The argument of periapsis is clustered around two values: $\sim40^{\circ}$
and $\sim180^{\circ}$, while the longitude of the ascending node
is clustered around $\sim0$ and $\sim150^{\circ}.$ The inclination
is also clustered around two values: $\sim40^{\circ}$ and $\sim140^{\circ}.$
Figure \ref{fig:Loss Fraction_inclined} presents the same as figure \ref{fig:Orbital-parameter-distribution-ejected}
for scenario (b). We find that the rate of the ejected particles in scenarion (b) is higher by an order of magnitude compared with scenario (a).

\section{Implications and Discussion}
\label{sec:Implications-and-Discussion}
Assuming that Planet Nine captured in the early stages of the solar
system lifetime, it must have interacted with the disk of planetesimal
around it. We chose the simplest structure of a debris disk that is
initially flat and circular. Our initial conditions are conservative
in the sense that disk, that is more inclined or eccentric would have
a lower specific angular momentum and therefore would tend to be more
influenced by Planet Nine. We have found that a spheroidal structure
(TAUS) is created around the Sun at $\sim1200{\rm AU}$ and is surrounded
by an inclined disk beyond $\sim1500{\rm AU}$.

\subsection{New structure in the solar-system}
\label{sub:New-structure-in}
The TAUS is a spheroidal structure. It does not represent a perfect
sphere because the particles' longitude of the ascending nodes is
not uniformly distributed. This implies that some parts of the sphere
are populated more than others. The number of particles in the TAUS
is not well constrained, as it depends on the initial conditions, number of
particles in the disk and their  surface density distribution.
Scenario (a) and (b) differ in the size of the TAUS; in scenario (a) TAUS end at $\sim1500\rm{AU}$
while in scenario (b) at $\sim3000\rm{AU}$.
We assumed the power law index of the density distribution  be $\gamma=-1$. Unlike the Oort cloud at
$\sim10^{5}{\rm AU}$ or even the Hills cloud at$\sim10^{4}{\rm AU}$
the TAUS is relatively close and perhaps could
be detected by a dedicated infrared survey. 

Figure \ref{fig:Loss Fraction} implies that in the early stage of
the solar system evolution, around $300{\rm Myr}$ after formation,
there was a spike in the rate of ejected objects from the disk into
the inner solar systems, by up to two orders of magnitude relative
to the current rate. This implies that there was an epoch during which
a large number of objects from the region of $\sim1000{\rm AU}$ crossed
the orbits of the inner planets, similarly to the ``heavy bombardment''
epoch described in the Nice model \citep{Levison2008}.

\subsection{New origin of comets}
\label{sub:New-origin-of}
Some TAUS objects are excited to high eccentricities due to resonance and secular
interaction with Planet Nine (see section \ref{sub:Ejected-particles}).
These objects enter the solar system with eccentricities close to
unity and therefore can be considered as comet candidates. Due to
the specific orientation of the TAUS, we expect the comets to originate
from $1100{\rm AU}\lesssim a_{{\rm comet}}\lesssim1500{\rm AU} (3000\rm{AU})$,
and have the distribution of the specific argument of periapsis, longitude of the ascending
node and inclination shown in Figures \ref{fig:Orbital-parameter-distribution-ejected} 
and \ref{fig:Comets_inclined}.
This is a prediction of our model that can be tested observationally.
For a total number of objects in the TAUS, $N$, one can estimate
the rate of these events per year, as 
\begin{equation}
\Gamma_{{\rm comet}}\sim10^{-2}-10^{-1}{\rm yr^{-1}}\left(\frac{N}{10^{12}}\right).\label{eq:comet rate}
\end{equation}

The Large Synoptic Survey Telescope (LSST) will survey 20,000 square
degrees of the sky about 2,000 times over 10 years \citep{LSSTScienceCollaboration2009}.
This survey is expected to discover $10,000$ new comets and potentially
shed light on the size distribution of long period comets, which is
currently unknown. The Starshot Breakthrough Initiatives could also
explore the solar system (http://breakthroughinitiatives.org/Concept/3/). 

The gravitational interaction of the comet candidates with the solar
systems planets was not explored here. An extensive study on this
interaction and its observational signatures will be studied elsewhere.

\section{Summary}
\label{sec:Summary}
We calculated the dynamical imprint of Planet Nine on the outer solar
system, assuming a flat circular initial debris disk out to a distance
of $7000{\rm AU}$. We showed that orbits with $a<3000{\rm AU}$ interacted
with Planet Nine over the lifetime of the solar system.

Our main conclusions are as follows:
\begin{itemize}
\item A spheroidal structure, TAUS, forms at $\sim1200{\rm AU}$ due to
MMRs with Planet Nine.
\item TAUS is not uniformly distributed.
\item The interaction of Planet Nine with the disk produces a qualitatively
different morphology than a fly-by interaction with a passing star.
\item Objects from TAUS that are excited to high eccentricities could become
comets. 
\end{itemize}
These predictions can be tested observationally with future surveys
such as LSST or the Starshot Breakthrough Initiative.

\section*{Acknowledgements}
We thank Matt Holman and Hagai B. Perets for helpful comments on
the manuscript. E.M. thanks Harvard's Institute for Theory and Computation
(ITC) for its kind hospitality during August 2016, when this paper
was initiated. E.M also acknowledges the European union career integration
grant \textquotedblleft GRAND,\textquotedblright{} and the Israel
science foundation excellence centre I-CORE grant 1829/12.

\bibliographystyle{mnras}

\appendix

\section{Disk distributions}
\label{sec:App}
In the appendix we present the test particles orbital properties for the three different regimes: (\mbox{i}) TAUS ;(\mbox{ii}) an inclined disk
and (\mbox{iii}) a warped disk relative to the ecliptic plane. Moreover we present the properties of the ejected particles from scenario (b).
\begin{figure*}
\includegraphics[width=17cm]{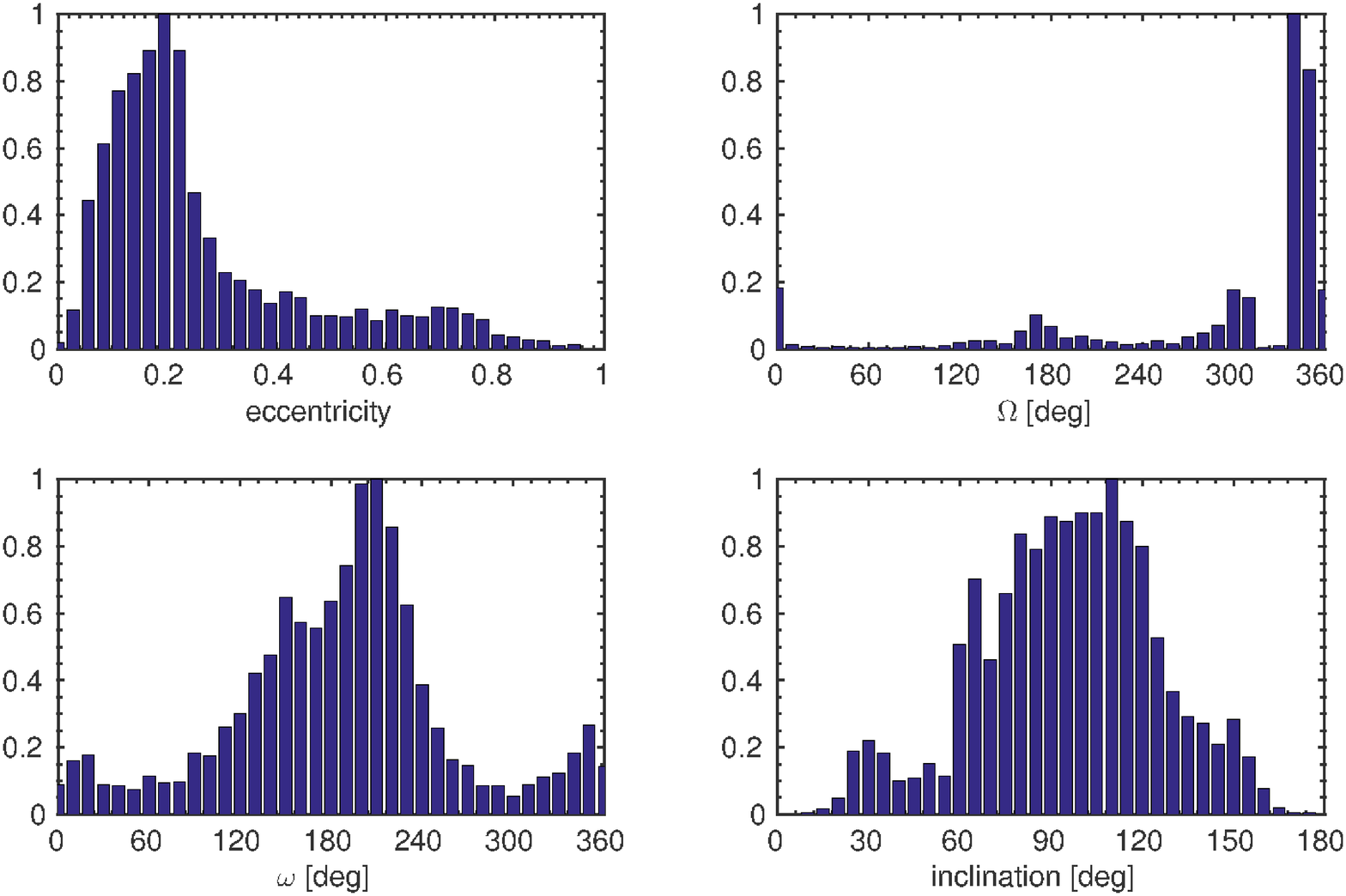}

\protect\caption{\label{fig:TAUS_ecc_inc_Omega_omega}Final distributions of the eccentricity,
longitude of ascending node and the argument of the periapsis for
TAUS. \textbf{Upper left:} The eccentricity distribution is concentrated
around $e\approx0.2$, with a tail of highly eccentric orbits (which
are stable on a $4\ {\rm Gyr}$ timescale). \textbf{Upper right:}
The longitude of ascending nodes show two characteristics: a narrow
distribution around $\Omega\approx350^{\circ}$ and a wide spread
tail. \textbf{Bottom left}: The argument of periapsis, $\omega,$
have a wider distribution around $\sim200^{\circ}.$ \textbf{Bottom
right:} The inclination distribution spans the range between $\sim30^{\circ}$and
$\sim160^{\circ}$, which gives the TAUS is spheroidal shape. All
distributions are normalized to the maximal value of unity.}
\end{figure*}

\begin{figure*}
\includegraphics[width=17cm]{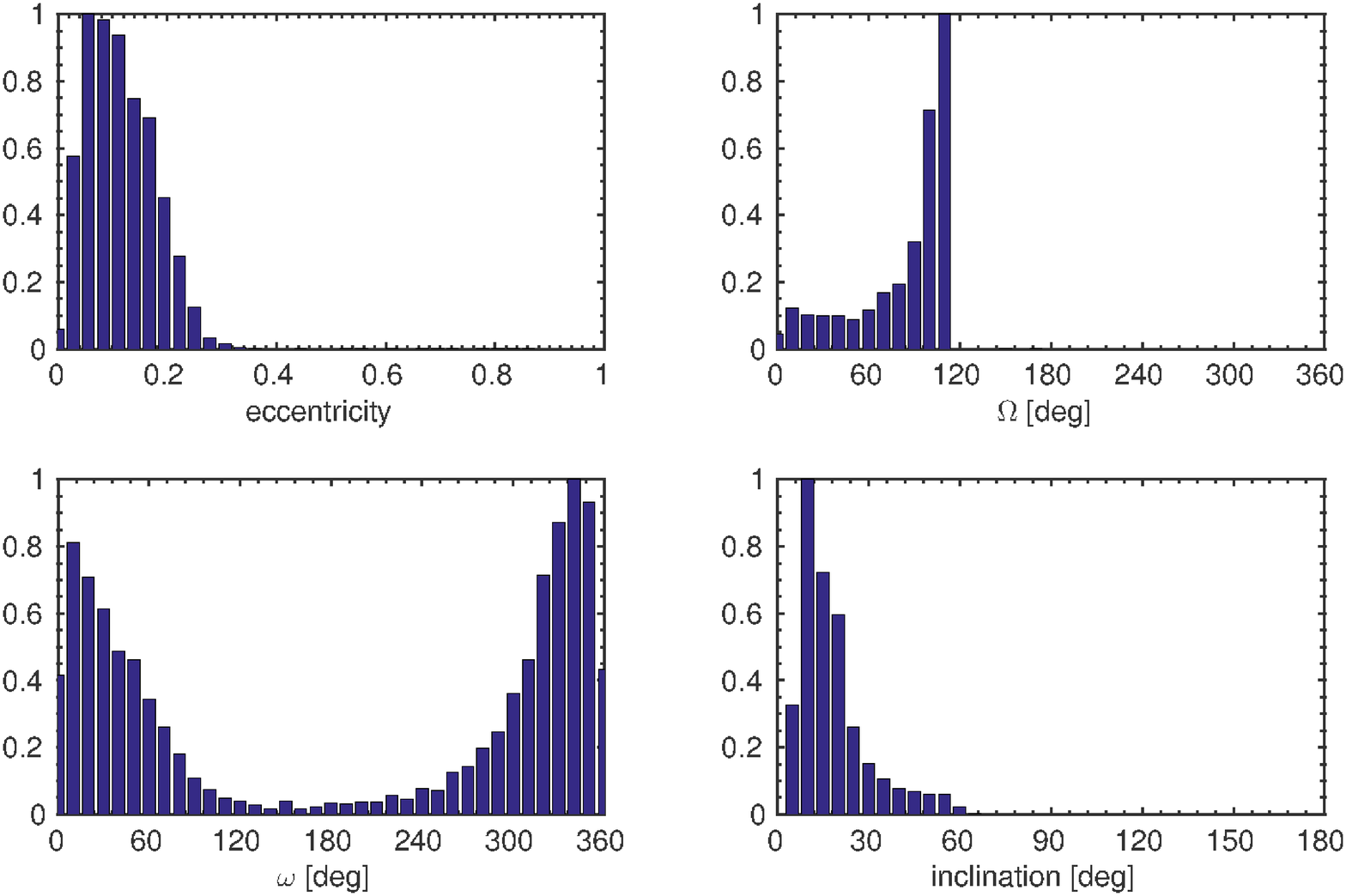}\protect\caption{\label{fig:InclinedDisk-Omega-omega-inc}Final distributions of the
eccentricity, longitude of ascending node and the argument of periapsis
for the inclined disk. \textbf{Upper left:} The eccentricity distribution
is concentrated around $e\approx0.1$. \textbf{Upper right:} The longitude
of the ascending nodes shows a sharp cutoff at $\sim110^{\circ}$.
\textbf{Bottom left:} The argument of the periapsis is clustered around
$\sim0$. \textbf{Bottom right:} The inclination distribution is narrow
with mean inclination of $18.4^{\circ}$.}
\end{figure*}

\begin{figure*}
\includegraphics[width=17cm]{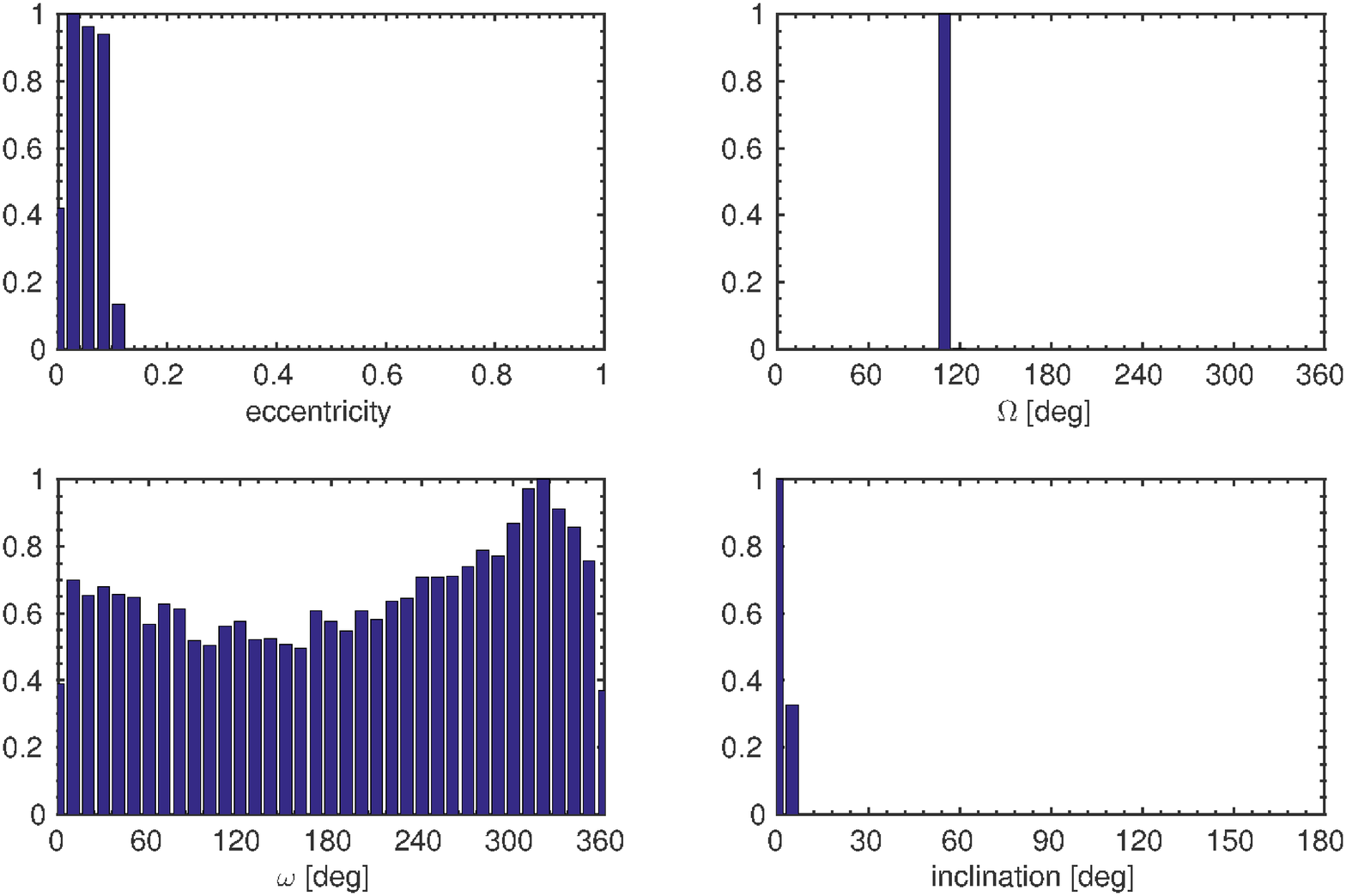}\protect\caption{\label{fig:Outer disk hist}Final distribution of the eccentricity,
longitude of ascending nodes and the argument of periapsis for the
outer disk. The distributions appear to follow the initial distributions.
This part of the disk did not have sufficient time to interact with
Planet Nine during the lifetime of the solar system.}
\end{figure*}

\begin{figure*}
\includegraphics[width=17cm]{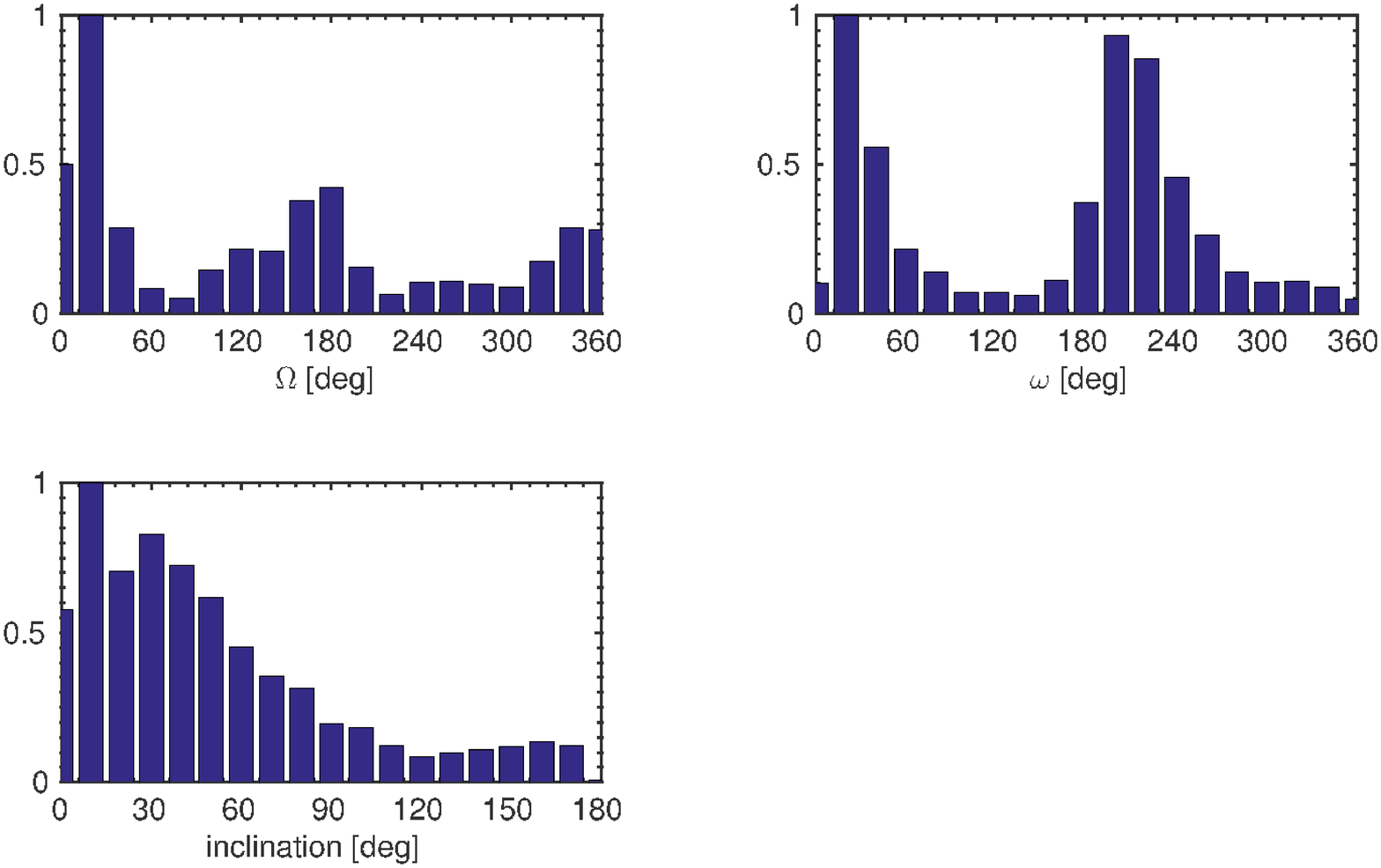}\protect\caption{\label{fig:Comets_inclined}Orbital parameter
distributions of the removed particles from the simulation from scenario (b), i.e. Sun grazing comets candidates. The argument
of periapsis has two sharp peaks around: $\sim30^{\circ}$ and $\sim200^{\circ}.$
The inclination distribution is wide. The longitude of ascending nodes
has two sharp bumps around: $\sim20$ and $\sim180^{\circ}$. All distribution
are normalized a maximum value of unity.}
\end{figure*}

\begin{figure*}
\includegraphics[width=17cm]{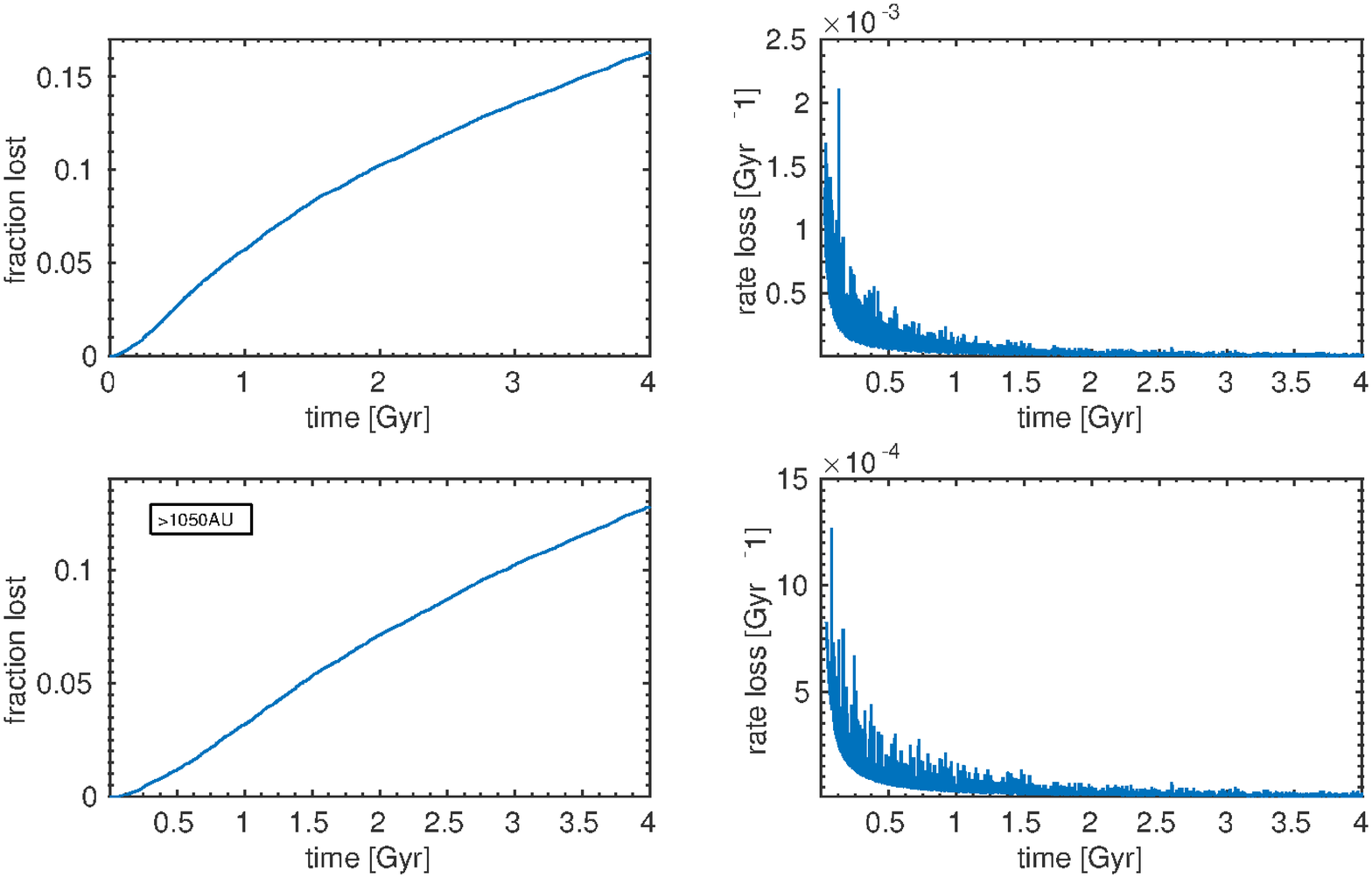}

\protect\caption{\label{fig:Loss Fraction_inclined}\textbf{Upper left:} Loss fraction from
the entire population of particles of scenario (b). During the integration time $\sim2\%$
of the particles were removed after entering the inner solar system.
\textbf{Upper right:} The rate of ejection as a function of time.
This panel is the time derivative of the left panel. \textbf{Bottom
left:} Loss fraction from the overall particles with initial sma greater
than $1050{\rm AU}$. \textbf{Bottom right:} The rate of ejection
as a function of time. We note that in this scenario the fraction loss and the loss rate is higher by an order of magnitude than scenario (a).}
\end{figure*}

\bsp	
\label{lastpage}
\end{document}